\documentstyle[10pt,epsf,epsfig,dp_delphititle]{dp_delphi}
\makeindex
\def\DpPaperGroup{EP}
\def\DpPaperRef{99--07}
\def\DpDate{20 January 1999}
\def\DpAuthors{DELPHI Collaboration}
\def\DpSubmit{(Submitted to Eur. Phys. J. C)}
\def\DpTitle{{ Measurement of the forward backward asymmetry \\
 of {\boldmath $c$} and {\boldmath $b$} quarks at the {\boldmath $Z$} pole \\
 using reconstructed {\boldmath $D$} mesons }}

\begin{document}
%%%%%%%%%%%%%%%%%%%%%%%%%% They are a problem with Coll.Sty ?
\makeatletter
%\input{dp_system:coll.sty}
% Collapse citation numbers to ranges.  Non-numeric and undefined labels
% are handled.  No sorting is done.  E.g., 1,3,2,3,4,5,foo,1,2,3,?,4,5
% gives 1,3,2-5,foo,1-3,?,4,5
\newcount\@tempcntc
\def\@citex[#1]#2{\if@filesw\immediate\write\@auxout{\string\citation{#2}}\fi
  \@tempcnta\z@\@tempcntb\m@ne\def\@citea{}\@cite{\@for\@citeb:=#2\do
    {\@ifundefined
       {b@\@citeb}{\@citeo\@tempcntb\m@ne\@citea\def\@citea{,}{\bf ?}\@warning
       {Citation `\@citeb' on page \thepage \space undefined}}%
    {\setbox\z@\hbox{\global\@tempcntc0\csname b@\@citeb\endcsname\relax}%
     \ifnum\@tempcntc=\z@ \@citeo\@tempcntb\m@ne
       \@citea\def\@citea{,}\hbox{\csname b@\@citeb\endcsname}%
     \else
      \advance\@tempcntb\@ne
      \ifnum\@tempcntb=\@tempcntc
      \else\advance\@tempcntb\m@ne\@citeo
      \@tempcnta\@tempcntc\@tempcntb\@tempcntc\fi\fi}}\@citeo}{#1}}
\def\@citeo{\ifnum\@tempcnta>\@tempcntb\else\@citea\def\@citea{,}%
  \ifnum\@tempcnta=\@tempcntb\the\@tempcnta\else
   {\advance\@tempcnta\@ne\ifnum\@tempcnta=\@tempcntb \else \def\@citea{--}\fi
    \advance\@tempcnta\m@ne\the\@tempcnta\@citea\the\@tempcntb}\fi\fi}
 
\makeatother
%%%%%%%%%%%%%%%%%%%%%%%%%% ??????????????????????????????????
% Generate the title page
\begin{titlepage}
\pagenumbering{roman}
\CERNpreprint{\DpPaperGroup}{\DpPaperRef} % Reference of the paper
\date{{\small\DpDate}} % Date of the paper
\title{\DpTitle} % Title of the paper
\address{\DpAuthors} % General name of the author(s)
\begin{shortabs} % Start the abstract
\noindent
%   abstract.tex
%
\newcommand {\Afbc}    {{\ifmmode  A_{FB}^{c}\else
                                   $A_{FB}^{c}$\fi}}
\newcommand {\Afbb}    {{\ifmmode  A_{FB}^{b}\else
                                   $A_{FB}^{b}$\fi}}
\newcommand {\SINEFF}  {{\ifmmode \sin^2\theta_{eff}^{lept}\else
                                  $\sin^2\theta_{eff}^{lept}$\fi}}

\noindent

A measurement of the forward--backward asymmetry of 
$e^{+}e^{-} \to c\bar{c}$ and 
$e^{+}e^{-} \to b\bar{b}$
on the $Z$ resonance is performed using about 3.5 million hadronic $Z$ decays
collected by the DELPHI detector at LEP in the years 1992 to 1995. The heavy
quark is tagged by the exclusive reconstruction of several $D$ meson decay
modes. The
forward--backward asymmetries for $c$ and $b$ quarks at the $Z$ resonance
are determined to be:

\[
    \renewcommand{\arraystretch}{1.6}
\begin{array}{rcr@{\!\!\;\,}l}
   \Afbc(\sqrt{s} = 91.235 \, {\rm GeV})
       &=& &0.0659\,\, \pm\,\, 0.0094\, (stat)\,\, \pm\,\, 0.0035\, (syst) \\
   \Afbb (\sqrt{s} = 91.235 \, {\rm GeV}) 
       &=& &0.0762\,\, \pm\,\, 0.0194\, (stat)\,\, \pm\,\, 0.0085\, (syst) \\
   \Afbc(\sqrt{s} = 89.434\, {\rm GeV})
       &=&-&0.0496\,\, \pm\,\, 0.0368\, (stat)\,\, \pm\,\, 0.0053\, (syst) \\
   \Afbb(\sqrt{s} = 89.434\, {\rm GeV})
       &=& &0.0567\,\, \pm\,\, 0.0756\, (stat)\,\, \pm\,\, 0.0117\, (syst) \\
   \Afbc(\sqrt{s} = 92.990\, {\rm GeV})
       &=& &0.1180\,\, \pm\,\, 0.0318\, (stat)\,\, \pm\,\, 0.0062\, (syst) \\
   \Afbb(\sqrt{s} = 92.990\, {\rm GeV})
       &=& &0.0882\,\, \pm\,\, 0.0633\, (stat)\,\, \pm\,\, 0.0122\, (syst) \\
\end{array}
\]
The combination of these results leads to an effective electroweak mixing
angle of:
$$\SINEFF = 0.2332 \pm 0.0016$$

%%% Local Variables: 
%%% mode: latex
%%% TeX-master: "paper"
%%% End: 
\end{shortabs}
\vfill
\begin{center}
\DpSubmit \ % Horrible hack to allow to have DpSubmit empty
%CD\DpComment \ \\
%CD\DpEMail \ \\
\end{center}
\vfill
\clearpage
\headsep 10.0pt
\addtolength{\textheight}{10mm}
\addtolength{\footskip}{-5mm}
\begingroup
% Commands to process the author names
%
\newcommand{\DpName}[2]{\hbox{#1$^{\ref{#2}}$},\hfill}
\newcommand{\DpNameTwo}[3]{\hbox{#1$^{\ref{#2},\ref{#3}}$},\hfill}
\newcommand{\DpNameThree}[4]{\hbox{#1$^{\ref{#2},\ref{#3},\ref{#4}}$},\hfill}
\newskip\Bigfill \Bigfill = 0pt plus 1000fill
\newcommand{\DpNameLast}[2]{\hbox{#1$^{\ref{#2}}$}\hspace{\Bigfill}}
%
%CD\small
\footnotesize
\noindent
\DpName{P.Abreu}{LIP}
\DpName{T.Adye}{RAL}
\DpName{P.Adzic}{DEMOKRITOS}
\DpName{Z.Albrecht}{KARLSRUHE}
\DpName{T.Alderweireld}{AIM}
\DpName{G.D.Alekseev}{JINR}
\DpName{R.Alemany}{VALENCIA}
\DpName{T.Allmendinger}{KARLSRUHE}
\DpName{P.P.Allport}{LIVERPOOL}
\DpName{S.Almehed}{LUND}
% ...
\normalsize
\endgroup
\titlefoot{Department of Physics and Astronomy, Iowa State
     University, Ames IA 50011-3160, USA
    \label{AMES}}
\titlefoot{Physics Department, Univ. Instelling Antwerpen,
     Universiteitsplein 1, BE-2610 Wilrijk, Belgium \\
     \indent~~and IIHE, ULB-VUB,
     Pleinlaan 2, BE-1050 Brussels, Belgium \\
     \indent~~and Facult\'e des Sciences,
     Univ. de l'Etat Mons, Av. Maistriau 19, BE-7000 Mons, Belgium
    \label{AIM}}
\titlefoot{Physics Laboratory, University of Athens, Solonos Str.
     104, GR-10680 Athens, Greece
    \label{ATHENS}}
\titlefoot{Department of Physics, University of Bergen,
     All\'egaten 55, NO-5007 Bergen, Norway
    \label{BERGEN}}
\titlefoot{Dipartimento di Fisica, Universit\`a di Bologna and INFN,
     Via Irnerio 46, IT-40126 Bologna, Italy
    \label{BOLOGNA}}
\titlefoot{Centro Brasileiro de Pesquisas F\'{\i}sicas, rua Xavier Sigaud 150,
     BR-22290 Rio de Janeiro, Brazil \\
     \indent~~and Depto. de F\'{\i}sica, Pont. Univ. Cat\'olica,
     C.P. 38071 BR-22453 Rio de Janeiro, Brazil \\
     \indent~~and Inst. de F\'{\i}sica, Univ. Estadual do Rio de Janeiro,
     rua S\~{a}o Francisco Xavier 524, Rio de Janeiro, Brazil
    \label{BRASIL}}
\titlefoot{Comenius University, Faculty of Mathematics and Physics,
     Mlynska Dolina, SK-84215 Bratislava, Slovakia
    \label{BRATISLAVA}}
\titlefoot{Coll\`ege de France, Lab. de Physique Corpusculaire, IN2P3-CNRS,
     FR-75231 Paris Cedex 05, France
    \label{CDF}}
\titlefoot{CERN, CH-1211 Geneva 23, Switzerland
    \label{CERN}}
\titlefoot{Institut de Recherches Subatomiques, IN2P3 - CNRS/ULP - BP20,
     FR-67037 Strasbourg Cedex, France
    \label{CRN}}
\titlefoot{Institute of Nuclear Physics, N.C.S.R. Demokritos,
     P.O. Box 60228, GR-15310 Athens, Greece
    \label{DEMOKRITOS}}
\titlefoot{FZU, Inst. of Phys. of the C.A.S. High Energy Physics Division,
     Na Slovance 2, CZ-180 40, Praha 8, Czech Republic
    \label{FZU}}
\titlefoot{Dipartimento di Fisica, Universit\`a di Genova and INFN,
     Via Dodecaneso 33, IT-16146 Genova, Italy
    \label{GENOVA}}
\titlefoot{Institut des Sciences Nucl\'eaires, IN2P3-CNRS, Universit\'e
     de Grenoble 1, FR-38026 Grenoble Cedex, France
    \label{GRENOBLE}}
\titlefoot{Helsinki Institute of Physics, HIP,
     P.O. Box 9, FI-00014 Helsinki, Finland
    \label{HELSINKI}}
\titlefoot{Joint Institute for Nuclear Research, Dubna, Head Post
     Office, P.O. Box 79, RU-101 000 Moscow, Russian Federation
    \label{JINR}}
\titlefoot{Institut f\"ur Experimentelle Kernphysik,
     Universit\"at Karlsruhe, Postfach 6980, DE-76128 Karlsruhe,
     Germany
    \label{KARLSRUHE}}
\titlefoot{Institute of Nuclear Physics and University of Mining and Metalurgy,
     Ul. Kawiory 26a, PL-30055 Krakow, Poland
    \label{KRAKOW}}
\titlefoot{Universit\'e de Paris-Sud, Lab. de l'Acc\'el\'erateur
     Lin\'eaire, IN2P3-CNRS, B\^{a}t. 200, FR-91405 Orsay Cedex, France
    \label{LAL}}
\titlefoot{School of Physics and Chemistry, University of Lancaster,
     Lancaster LA1 4YB, UK
    \label{LANCASTER}}
\titlefoot{LIP, IST, FCUL - Av. Elias Garcia, 14-$1^{o}$,
     PT-1000 Lisboa Codex, Portugal
    \label{LIP}}
\titlefoot{Department of Physics, University of Liverpool, P.O.
     Box 147, Liverpool L69 3BX, UK
    \label{LIVERPOOL}}
\titlefoot{LPNHE, IN2P3-CNRS, Univ.~Paris VI et VII, Tour 33 (RdC),
     4 place Jussieu, FR-75252 Paris Cedex 05, France
    \label{LPNHE}}
\titlefoot{Department of Physics, University of Lund,
     S\"olvegatan 14, SE-223 63 Lund, Sweden
    \label{LUND}}
\titlefoot{Universit\'e Claude Bernard de Lyon, IPNL, IN2P3-CNRS,
     FR-69622 Villeurbanne Cedex, France
    \label{LYON}}
\titlefoot{Univ. d'Aix - Marseille II - CPP, IN2P3-CNRS,
     FR-13288 Marseille Cedex 09, France
    \label{MARSEILLE}}
\titlefoot{Dipartimento di Fisica, Universit\`a di Milano and INFN,
     Via Celoria 16, IT-20133 Milan, Italy
    \label{MILANO}}
\titlefoot{Niels Bohr Institute, Blegdamsvej 17,
     DK-2100 Copenhagen {\O}, Denmark
    \label{NBI}}
\titlefoot{NC, Nuclear Centre of MFF, Charles University, Areal MFF,
     V Holesovickach 2, CZ-180 00, Praha 8, Czech Republic
    \label{NC}}
\titlefoot{NIKHEF, Postbus 41882, NL-1009 DB
     Amsterdam, The Netherlands
    \label{NIKHEF}}
\titlefoot{National Technical University, Physics Department,
     Zografou Campus, GR-15773 Athens, Greece
    \label{NTU-ATHENS}}
\titlefoot{Physics Department, University of Oslo, Blindern,
     NO-1000 Oslo 3, Norway
    \label{OSLO}}
\titlefoot{Dpto. Fisica, Univ. Oviedo, Avda. Calvo Sotelo
     s/n, ES-33007 Oviedo, Spain
    \label{OVIEDO}}
\titlefoot{Department of Physics, University of Oxford,
     Keble Road, Oxford OX1 3RH, UK
    \label{OXFORD}}
\titlefoot{Dipartimento di Fisica, Universit\`a di Padova and
     INFN, Via Marzolo 8, IT-35131 Padua, Italy
    \label{PADOVA}}
\titlefoot{Rutherford Appleton Laboratory, Chilton, Didcot
     OX11 OQX, UK
    \label{RAL}}
\titlefoot{Dipartimento di Fisica, Universit\`a di Roma II and
     INFN, Tor Vergata, IT-00173 Rome, Italy
    \label{ROMA2}}
\titlefoot{Dipartimento di Fisica, Universit\`a di Roma III and
     INFN, Via della Vasca Navale 84, IT-00146 Rome, Italy
    \label{ROMA3}}
\titlefoot{DAPNIA/Service de Physique des Particules,
     CEA-Saclay, FR-91191 Gif-sur-Yvette Cedex, France
    \label{SACLAY}}
\titlefoot{Instituto de Fisica de Cantabria (CSIC-UC), Avda.
     los Castros s/n, ES-39006 Santander, Spain
    \label{SANTANDER}}
\titlefoot{Dipartimento di Fisica, Universit\`a degli Studi di Roma
     La Sapienza, Piazzale Aldo Moro 2, IT-00185 Rome, Italy
    \label{SAPIENZA}}
\titlefoot{Inst. for High Energy Physics, Serpukov
     P.O. Box 35, Protvino, (Moscow Region), Russian Federation
    \label{SERPUKHOV}}
\titlefoot{J. Stefan Institute, Jamova 39, SI-1000 Ljubljana, Slovenia
     and Laboratory for Astroparticle Physics,\\
     \indent~~Nova Gorica Polytechnic, Kostanjeviska 16a, SI-5000 Nova Gorica, Slovenia, \\
     \indent~~and Department of Physics, University of Ljubljana,
     SI-1000 Ljubljana, Slovenia
    \label{SLOVENIJA}}
\titlefoot{Fysikum, Stockholm University,
     Box 6730, SE-113 85 Stockholm, Sweden
    \label{STOCKHOLM}}
\titlefoot{Dipartimento di Fisica Sperimentale, Universit\`a di
     Torino and INFN, Via P. Giuria 1, IT-10125 Turin, Italy
    \label{TORINO}}
\titlefoot{Dipartimento di Fisica, Universit\`a di Trieste and
     INFN, Via A. Valerio 2, IT-34127 Trieste, Italy \\
     \indent~~and Istituto di Fisica, Universit\`a di Udine,
     IT-33100 Udine, Italy
    \label{TU}}
\titlefoot{Univ. Federal do Rio de Janeiro, C.P. 68528
     Cidade Univ., Ilha do Fund\~ao
     BR-21945-970 Rio de Janeiro, Brazil
    \label{UFRJ}}
\titlefoot{Department of Radiation Sciences, University of
     Uppsala, P.O. Box 535, SE-751 21 Uppsala, Sweden
    \label{UPPSALA}}
\titlefoot{IFIC, Valencia-CSIC, and D.F.A.M.N., U. de Valencia,
     Avda. Dr. Moliner 50, ES-46100 Burjassot (Valencia), Spain
    \label{VALENCIA}}
\titlefoot{Institut f\"ur Hochenergiephysik, \"Osterr. Akad.
     d. Wissensch., Nikolsdorfergasse 18, AT-1050 Vienna, Austria
    \label{VIENNA}}
\titlefoot{Inst. Nuclear Studies and University of Warsaw, Ul.
     Hoza 69, PL-00681 Warsaw, Poland
    \label{WARSZAWA}}
\titlefoot{Fachbereich Physik, University of Wuppertal, Postfach
     100 127, DE-42097 Wuppertal, Germany
    \label{WUPPERTAL}}
\titlefoot{On leave of absence from IHEP Serpukhov
    \label{MILAN-SERPOU}}
\titlefoot{Now at University of Florida
    \label{FLORIDA}}
\addtolength{\textheight}{-10mm}
\addtolength{\footskip}{5mm}
\clearpage
\headsep 30.0pt
\end{titlepage}
%%%%%%%%%%%%%%%%%%%%%%%%%
%
% Change for the document body
%%%CD\pagestyle{heading} % for page numbering
\pagenumbering{arabic} % page numbering in number
\setcounter{footnote}{0} %
\large
%   document.tex
%
\newfont{\tenfont}{cmr5 scaled 500}
\newcommand{\db}     {{\ifmmode {D}\hspace*{-0.67em}{\mbox{\raisebox{1.79ex}[0cm][0cm]{{\tenfont (}\rule[0.15mm]{1.4mm}{0.09mm}{\tenfont )}}}}
                        \else ${D}\hspace*{-0.67em}{\mbox{\raisebox{1.79ex}[0cm][0cm]{{\tenfont (}\rule[0.15mm]{1.4mm}{0.09mm}{\tenfont )}}}}$\fi}}
\newcommand{\dnb}     {{\ifmmode {D^0}\hspace*{-1.07em}{\mbox{\raisebox{1.79ex}[0cm][0cm]{{\tenfont (}\rule[0.15mm]{1.4mm}{0.09mm}{\tenfont )}}}}\,\,
                        \else ${D^0}\hspace*{-1.07em}{\mbox{\raisebox{1.79ex}[0cm][0cm]{{\tenfont (}\rule[0.15mm]{1.4mm}{0.09mm}{\tenfont )}}}}$\,\,\fi}}
\newcommand{\dnsb}     {{\ifmmode {D^{\ast 0}}\hspace*{-1.44em}{\mbox{\raisebox{1.79ex}[0cm][0cm]{{\tenfont (}\rule[0.15mm]{1.4mm}{0.09mm}{\tenfont )}}}}\,\,\,\,
                        \else ${D^{\ast 0}}\hspace*{-1.44em}{\mbox{\raisebox{1.79ex}[0cm][0cm]{{\tenfont (}\rule[0.15mm]{1.4mm}{0.09mm}{\tenfont )}}}}$\,\,\,\,\fi}}
\newcommand{\ds}      {{\ifmmode    D^{\ast +}
                         \else      $D^{\ast +}$ \fi}}
\newcommand{\dss}      {{\ifmmode    D^{+}_s
                         \else      $D^{+}_s$ \fi}}
\newcommand{\lc}      {{\ifmmode    \Lambda^{+}_c
                         \else      $\Lambda^{+}_c$ \fi}}
\newcommand{\dn}      {{\ifmmode    D^{0}
                         \else      $D^{0}$ \fi}}
\newcommand{\dnpi}      {{\ifmmode    D^{0} \pi^{+}
                           \else      $D^{0} \pi^{+}$ \fi}}
\newcommand{\dpl}     {{\ifmmode    D^{+}
                         \else      $D^{+}$ \fi}}

\newcommand{\klnpis}     {{\ifmmode    (K^- l^{+} (\nu_l)) \pi^+     
                         \else      $(K^- l^{+} (\nu_l)) \pi^+$ \fi}}
\newcommand{\kenpis}      {{\ifmmode    (K^- e^{+} (\nu_e))\pi^+ 
                         \else      $(K^- e^{+} (\nu_e))\pi^+$ \fi}}
\newcommand{\kunpis}      {{\ifmmode    (K^- \mu^{+} (\nu_\mu)) \pi^+
                         \else      $(K^- \mu^{+} (\nu_\mu)) \pi^+$ \fi}}
\newcommand{\kpipis}      {{\ifmmode    (K^- \pi^+) \pi^+    
                         \else      $(K^- \pi^+)\pi^+$ \fi}}
\newcommand{\kpipi}      {{\ifmmode    K^- \pi^+ \pi^+     
                         \else      $K^- \pi^+ \pi^+$ \fi}}
\newcommand{\kpi}      {{\ifmmode    K^- \pi^+     
                         \else      $K^- \pi^+$ \fi}}
\newcommand{\ktpipis}  {{\ifmmode    (K^- \pi^+ \pi^0)\pi^+    
                         \else      $(K^- \pi^+ \pi^0)\pi^+$ \fi}}
\newcommand{\kfpipis}  {{\ifmmode    (K^- \pi^+ \pi^- \pi^+)  \pi^+    
                         \else      $(K^- \pi^+ \pi^- \pi^+)\pi^+$ \fi}}
\newcommand{\kpipin}   {{\ifmmode    K^- \pi^+ (\pi^0)   
                         \else      $K^- \pi^+ (\pi^0)$ \fi}}
\newcommand{\kpipinpis} {{\ifmmode    (K^- \pi^+ (\pi^0)) \pi^+  
                         \else      $(K^- \pi^+ (\pi^0)) \pi^+$  \fi}}
\newcommand{\ksk}     {{\ifmmode   K^{\ast 0)} K^+
                         \else      $K^{\ast 0)} K^+$  \fi}}
\newcommand{\kpik}    {{\ifmmode   (K^- \pi^+ ) K^+
                         \else      $(K^- \pi^+ ) K^+$  \fi}}
\newcommand{\kkpi}    {{\ifmmode   (K^+ K^-) \pi^+ 
                         \else      $(K^+ K^-) \pi^+  $  \fi}}
\newcommand{\phipi}   {{\ifmmode   \phi \pi^+
                         \else      $\phi \pi^+$  \fi}}
\newcommand{\pkpi}    {{\ifmmode   P K^- \pi^+
                         \else      $P K^- \pi^+$  \fi}}
\newcommand{\dstokpi} {{\ifmmode    D^{\ast +} \to (K^- \pi^+) \pi^+
                         \else      $D^{\ast +} \to (K^- \pi^+) \pi^+$  \fi}}
\newcommand{\dstokfpi} {{\ifmmode    D^{\ast +} \to (K^- \pi^+ \pi^- \pi^+)  \pi^+
                         \else      $D^{\ast +} \to (K^- \pi^+ \pi^- \pi^+)  \pi^+$  \fi}}
\newcommand{\dstoktpi} {{\ifmmode    D^{\ast +} \to (K^- \pi^+ \pi^0)\pi^+
                         \else      $D^{\ast +} \to (K^- \pi^+ \pi^0)\pi^+$  \fi}}
\newcommand{\dstokln} {{\ifmmode    D^{\ast +} \to (K^- l^{+} (\nu_l)) \pi^+
                         \else      $D^{\ast +} \to (K^- l^{+} (\nu_l)) \pi^+$  \fi}}
\newcommand{\dstoken} {{\ifmmode    D^{\ast +} \to (K^- e^{+} (\nu_e))\pi^+
                         \else      $D^{\ast +} \to (K^- e^{+} (\nu_e))\pi^+$  \fi}}
\newcommand{\dstokun} {{\ifmmode    D^{\ast +} \to (K^- \mu^{+} (\nu_\mu)) \pi^+
                         \else      $D^{\ast +} \to (K^- \mu^{+} (\nu_\mu)) \pi^+$  \fi}}
\newcommand{\dstokpipin} {{\ifmmode    D^{\ast +} \to (K^- \pi^+ (\pi^0)) \pi^+
                         \else      $D^{\ast +} \to (K^- \pi^+ (\pi^0)) \pi^+$  \fi}}
\newcommand{\dntokpin} {{\ifmmode    D^{\ast +} \to K^- \pi^+ (\pi^0)
                         \else      $D^{\ast +} \to K^- \pi^+ (\pi^0)$  \fi}}
\newcommand{\dntokpi} {{\ifmmode    D^{\ast +} \to K^- \pi^+
                         \else      $D^{\ast +} \to K^- \pi^+$  \fi}}

\newcommand{\dnto} {{\ifmmode    D^{\ast +} \to\else 
                                 $D^{\ast +} \to $\fi}}
\newcommand {\Afb}    {{\ifmmode   A_{FB}\else
                                   $A_{FB}$\fi}}
\newcommand {\Afbc}    {{\ifmmode  A_{FB}^{c}\else
                                    $A_{FB}^{c}$\fi}}
\newcommand {\Afbb}    {{\ifmmode  A_{FB}^{b}\else
                                   $A_{FB}^{b}$\fi}}
\newcommand {\Afbnc}    {{\ifmmode  A_{FB}^{0,c}\else
                                    $A_{FB}^{0,c}$\fi}}
\newcommand {\Afbnb}    {{\ifmmode  A_{FB}^{0,b}\else
                                   $ A_{FB}^{0,b}$\fi}}
\newcommand {\Afbf}    {{\ifmmode  A_{FB}^{f}\else
                                   $A_{FB}^{f}$\fi}}
\newcommand {\Afbnf}    {{\ifmmode  A_{FB}^{0,f}\else
                                    $A_{FB}^{0,f}$\fi}}
\newcommand {\Afbq}    {{\ifmmode  A_{FB}^{q}\else
                                   $A_{FB}^{q}$\fi}}
\newcommand {\pisl}    {{\ifmmode  \pi_{slow}\else
                                   $\pi_{slow}$\fi}}
\newcommand {\rbc}     {{\ifmmode  \rm{R_{b} P_{b \to D} / R_{c} P_{c \to
                                 D}}\else                            
                                   $\rm{R_{b} P_{b \to D} / R_{c} P_{c \to D}}\fi}}
\newcommand {\taub}    {{\ifmmode  \tau_{b}\else
                                   $\tau_{b}$\fi}}
\newcommand {\chieff}  {{\ifmmode  \chi_{eff}\else
                                   $\chi_{eff}$\fi}}
\newcommand {\avxec}   {{\ifmmode  \langle X_{E} \rangle_{c}\else
                                   $\langle X_{E} \rangle_{c}$\fi}}
\newcommand {\avxeb}   {{\ifmmode  \langle X_{E} \rangle_{b}\else
                                   $\langle X_{E} \rangle_{b}$\fi}}
\newcommand {\avxebd}  {{\ifmmode  \langle X_{E} \rangle_{B \to D}\else
                                   $\langle X_{E} \rangle_{B \to D}$\fi}}
\newcommand {\gcc}     {{\ifmmode  g \to c\bar{c}\else
                                   $g \to c\bar{c}$\fi}}
\newcommand {\ccbar}     {{\ifmmode  c\bar{c}\else
                                     $c\bar{c}$\fi}}
\newcommand {\bbbar}     {{\ifmmode  b\bar{b}\else
                                     $b\bar{b}$\fi}}
\newcommand {\mpm}     {{\ifmmode  \pm\else
                                   $\pm$\fi}}
\newcommand {\mmp}     {{\ifmmode  \mp\else
                                   $\mp$\fi}}

\newcommand {\COSTH}  {{\ifmmode  \cos\theta\else
                                  $\cos\theta$\fi}}
\newcommand {\COSHEL} {{\ifmmode  \cos\theta_h\else
                                  $\cos\theta_h$\fi}}
\newcommand {\SINWB}  {{\ifmmode  \sin^2\theta_W\else
                                  $\sin^2\theta_W$\fi}}
\newcommand {\SINEFF}  {{\ifmmode \sin^2\theta_{eff}^{lept}\else
                                  $\sin^2\theta_{eff}^{lept}$\fi}}
\newcommand {\rphi}  {{\ifmmode R\phi\else
                                $R\phi$\fi}}
\newcommand {\dedx}  {{\ifmmode dE / dx\else
                                $dE / dx$\fi}}
\newcommand {\bbnm}  {{\ifmmode B^0 -\bar{B}^0\else
                                $B^0 -\bar{B}^0$\fi}}
\newcommand {\gevc}  {{\ifmmode  GeV/c\else 
                                 $Gev/c$\fi}}
\newcommand {\gevcc}  {{\ifmmode  GeV/c^2\else 
                                  $Gev/c^2$\fi}}
\newcommand {\gev}  {{\ifmmode  GeV\else 
                                $Gev$\fi}}
%\newcommand {}  {{\ifmmode 
%                          \else    $$  \fi}}

%========================================================================%
\section{Introduction}
%=========================================================================%

The cross--section for the process $e^{+}e^{-} \to Z \to
f\bar{f}$ for the fermion $f$ as a function of its polar angle $\theta$ with
respect to the direction of the $e^{-}$ can be expressed as:
$$
\frac{d\sigma}{d \cos\theta} \propto 1 + \frac{8}{3}\, \Afbf \,\cos\theta + \cos^2\theta\,.
$$
The term proportional to $\cos\theta$ generates a forward--backward asymmetry
\Afbf which results from the interference of the vector ($v$) and axial
vector ($a$) couplings of the initial and final state fermions of the $Z$
boson. The improved Born level asymmetry using the effective couplings
for pure $Z$ exchange is given by:
$$
\Afbf = \frac{3}{4} \,\,
         \frac{2\, \bar{v}_{\rm e} \,\bar{a}_{\rm e}}{\bar{v}^2_{\rm e} + \bar{a}^2_{\rm e} }\,\,
         \frac{2\, \bar{v}_f\, \bar{a}_f}{\bar{v}^2_f + \bar{a}^2_f }
$$
The measurement of the forward--backward asymmetry for the different
fermions in the final state can thus be used to measure $v/a$, and hence to 
determine the electroweak mixing angle \SINEFF.

In this analysis, the forward--backward
asymmetries for the processes $e^+e^-\rightarrow c\bar{c}$ and
$e^+e^-\rightarrow b\bar{b}$ at the $Z$ resonance are measured using
reconstructed $D$ mesons in the 
modes\footnote{Throughout the paper charge-conjugated states are 
included implicitly.}: \\

\begin{tabular}{llll}
 \ds  &$\to$  \dnpi                             &&   \\
      &     \phantom{$\to$}\,\,$\to$ \kpipis    &&   \\
      &     \phantom{$\to$}\,\,$\to$ \kfpipis   &&   \\
      &     \phantom{$\to$}\,\,$\to$ \kpipinpis &&
             with and without $\pi^0$ reconstruction \\
      &     \phantom{$\to$}\,\,$\to$ \kunpis    &&   \\
      &     \phantom{$\to$}\,\,$\to$ \kenpis    &&   \\
 \dn  &$\to$  \kpi                              &&   \\
 \dn  &$\to$  \kpipin                           &&   

                         without $\pi^0$ reconstruction   \\
 \dpl &$\to$  \kpipi                            &&   \\
\end{tabular} \\

The $D$ meson contains a charm quark and therefore provides a clean signature
of a $c\bar{c}$ event or a decay of a heavy $b$--hadron in a $b\bar{b}$
event. In both cases the charge state\footnote{For the $D^0$ 
%LLLLLLLLLL    I put in 'minus'
the charge state is defined as minus the charge of its decay kaon.} of the $D$
is directly correlated to the charge
of the primary quark. 

Particle identification in the DELPHI detector is provided by  ring imaging 
Cherenkov
counters (RICH), the specific energy loss $dE/dx$ in the Time Projection 
Chamber (TPC), and electron and muon
identification (see section 2); together with $\pi^0$ reconstruction, it is 
used to identify the $D$ decay
products and to reduce the combinatorial background.

It is necessary to distinguish between the contributions
of $D$ mesons from $c$ and $b$ quark events in order to determine the individual 
forward--backward asymmetries. In this
analysis a simultaneous fit is performed in terms of the scaled energy 
$X_E = 2 E_D / \sqrt s$ (where $E_D$ is the energy of the $D$)
and the $b$-tagging probability variable (see section \ref{measurement}). The
forward--backward asymmetry is then extracted from the distribution of the 
cosine of the polar angle of the thrust axis signed by the charge state of 
the $D$ meson.

In this paper an update of previous DELPHI results \cite{markus} is presented.
All \mbox{LEP 1} data collected by the DELPHI detector in the years 1992 to
1995 are used. The high quality of the data through the final LEP 1
reprocessing in combination with improved reconstruction and $b$-tagging
techniques \cite{delphi_rb} result in a significant gain in statistical
precision.

%=========================================================================%
\section{The DELPHI Detector}
%=========================================================================%

The DELPHI detector consists of several independent devices
for tracking, calorimetry and lepton and hadron identification.
The components relevant for this analysis 
will be briefly described
in the following. A detailed description of the whole apparatus 
and its performance can be found in~\cite{performance}.

In the barrel region the innermost component is the vertex detector (VD)
near to the LEP beam pipe. 
The VD 
%LEP I version of the
%detector 
consists of three concentric layers
(closer, inner and outer) of silicon microstrip detectors.
Since 1994 the VD provides $R\phi$ and $z$ 
%LLLLLLLLLLLLLL I've put in a footnote
information\footnote{In the DELPHI coordinate system, $z$ is along the
electron beam
direction, $\phi$ and $R$ are the azimuthal angle and radius in the $xy$ plane,
and $\theta$ is the polar angle with respect to the $z$ axis.} 
in the closer and outer layer
and has an extended polar angle
coverage of $25^\circ < \theta < 155^\circ$ (closer layer).
For polar angles of $44^\circ < \theta < 136^\circ$, a particle crosses all
three layers of the VD. With an intrinsic
$R\phi$ resolution of $7.6 \, {\rm \mu m}$~\cite{performance}, the VD is
the main component for  reconstructing secondary vertices of heavy hadron
decays.

The inner detector (ID) is outside the VD and consisted
%The VD is followed by the inner detector
%(ID). It consisted 
of a jet-chamber to perform a precise $R \phi$ measurement,
and five cylindrical MWPC layers. In 1995 the MWPC layers were replaced by
five layers of straw tube detectors.

The ID is followed by the TPC, the main tracking device in DELPHI.
It covers polar angles between $21^\circ < \theta < 159^\circ$
with a single point resolution for charged particles of approximately
$250 \, \mu$m in $R \phi$ and $880 \, \mu$m in $z$~\cite{performance}.
The analysis  of the pulse heights of the signals of up to 192 sense wires
allows the determination of the specific energy loss, \dedx , of charged
particles which can
be used for particle identification (see section \ref{sec_rek}).

The barrel ring imaging Cherenkov counter is  behind the TPC. Its
gas and liquid radiators allow particle identification for pions, kaons
and protons over almost the whole
momentum range (see section \ref{sec_rek}).

The outer detector (OD) is mounted behind the RICH to give additional tracking
information. It improves 
significantly the momentum resolution due to its large distance from the
interaction point. Five layers of drift cells cover polar angles between 
$42^\circ < \theta < 138^\circ$ and provide $R\phi$ and $z$ information.

The barrel electromagnetic calorimeter (HPC) is between the OD and the
superconducting coil and covers polar angles between
$42^\circ < \theta < 138^\circ$. It is a gas-sampling device which provides
complete three-dimensional charge information in the same way as a time
projection chamber. The excellent granularity allows good separation
between close particles in three dimensions. This permits good electron
identification even inside jets and direct identification of
$\pi^0 \to \gamma \gamma$ decays.

In the forward region,  tracking is performed by two planar drift chambers
(FCA and FCB) with a polar angle covering of $11^\circ < \theta < 33^\circ$
(FCA) and $11^\circ < \theta < 36.5^\circ$ (FCB). Their resolutions transverse
to the beam axis 
%in $x$
%and $y$ 
are $270 \, \mu$m (FCA) and $150 \, \mu$m (FCB) respectively.

For muon identification the DELPHI detector is surrounded by layers of
drift chambers. They cover $52^\circ < \theta < 128^\circ$ with a resolution
of $4$\,mm in $R\phi$ and $2.5$\,cm in $z$ in the barrel region
and $9^\circ < \theta < 43^\circ$ 
with a resolution of $1$\,mm in the forward region.

%=========================================================================%
\section{Hadronic event selection}
%=========================================================================%

Charged particles are selected as follows.
The momentum is required to be between 0.4~GeV/$c$ and 50~GeV/$c$, 
the relative error on the momentum measurement less than 100\,\%,
the polar angle relative to the beam axis between $20^{\circ}$ and 
$160^{\circ}$,
the length of tracks in the TPC  larger than 30\,cm,
the projection of the impact
parameter relative to the interaction point  less than 4\,cm
in the plane transverse to the beam direction, and the distance 
along the beam direction to the interaction point less than 10\,cm.
 
Hadronic events are selected by requiring five or more charged particles
and a total energy in charged particles larger than 12\,\% of the collision
energy (assuming all charged particles to be pions).
A total of 3.5 million hadronic events is obtained from the 1992-1995 data,
at centre--of--mass energies within $\pm$ 2~GeV of the $Z$ resonance mass.
According to the simulation, the selection efficiency for hadronic $Z$ 
decays is $95.7\,\%$.
Table \ref{hadronic_numbers} shows
the number of selected hadronic events.
Remaining backgrounds from $\tau$ pairs and Bhabha events are found to be
negligible for this analysis.

A set of about 8.5 million simulated hadronic events for the years 1992 to
1995 is used. They are generated using JETSET 7.4 Parton Shower
model~\cite{jetset} in combination with the full
simulation of the DELPHI detector. The parameters of the generator are tuned
to the DELPHI data \cite{delphi_tuning}. 

For each event, the primary interaction vertex is
determined from the measured tracks, with
a constraint from the measured mean beam spot position. The removal of the
track with the largest $\chi^2$ (followed by a refit of the vertex)  is
repeated until either the $\chi ^2$ of each contributing track is
less than 3 or less than three charged particle tracks are left.
%or at least two charged particle tracks are left. 
All track parameters are recalculated
after a helix extrapolation to this vertex position.
The resolution of tracks measured only by the forward tracking chambers
is improved by a track refit using the primary vertex. Forward tracks having a 
$\chi^2$ in the refit larger than 100 are removed from the analysis.

%\begin{table}[ht]
% \begin{center}
%  \begin{tabular}{|c||r|r|r||r|c|} \hline
%Year  & \multicolumn{3}{c||}{Data events}    & \multicolumn{2}{c|}{Simulation} \\
%      &$91.235$\;GeV& $89.434$\;GeV&$92.990$\;GeV  &  \multicolumn{1}{c|}{events} 
%      & efficiency\\ \hline\hline 
%1992  & 703859~  &---  ~~~~  &--- ~~~~~  & 2003142     & 95.7 \% \\ \hline
%1993  & 475151~  &  97623~~~ & 134240~~~ & 1893139     & 95.6 \% \\ \hline
%1994  & 1386191~ &---  ~~~~  &---  ~~~~~ & 3551362     & 95.7 \% \\ \hline
%1995  & 458700~  &  84763~~~ & 131637~~~ & 1126557     & 95.8 \% \\ \hline \hline
%92-95 & 3023901~ & 182386~~~ & 265877~~~ & 8574200     & 95.7 \% \\ \hline
%   \end{tabular}  
%  \end{center}
% \caption{The number of selected hadronic events for data and simulation
%          and the selection efficiency taken from the simulation.}
% \label{hadronic_numbers}
%\end{table}
\begin{table}[ht]
 \begin{center}
  \begin{tabular}{|c||r|r|r||c|} \hline
Year  & \multicolumn{3}{c||}{Data events}    & Simulation \\
      &91.235\,GeV& 89.434\,GeV&92.990\,GeV  & events    \\ \hline\hline 
1992  & 703859~  &---  ~~~~  &--- ~~~~~  & 2003142     \\ \hline
1993  & 475151~  &  97623~~~ & 134240~~~ & 1893139     \\ \hline
1994  & 1386191~ &---  ~~~~  &---  ~~~~~ & 3551362     \\ \hline
1995  & 458700~  &  84763~~~ & 131637~~~ & 1126557     \\ \hline \hline
92-95 & 3023901~ & 182386~~~ & 265877~~~ & 8574200     \\ \hline
   \end{tabular}  
  \end{center}
 \caption{The number of selected hadronic events for data and simulation.}
 \label{hadronic_numbers}
\end{table}

%=========================================================================%
\section{\label{sec_rek} Reconstruction of charmed mesons}
%=========================================================================%

Reconstructed $D$ mesons are used as a signature for
$c\bar{c}$ and $b\bar{b}$ events and are identified through their
decay products. The $D$ mesons are reconstructed in nine different
decay modes (see table \ref{xedlcuts}). In the following a brief description of the selection
criteria for the $D$ candidates is given.

For all decay modes the selection of candidates is performed in a similar
way. A number of charged particles 
(corresponding to the multiplicity of the specific $D^{0/+}$ decay mode) 
with momentum p larger than 1 GeV/$c$ 
are combined, 
requiring the total charge to be zero in case of the $D^0$ and one in
case of the $D^+$ decay. The invariant mass m$_D$ of the $D^{0/+}$ candidate
is calculated, assuming one of the particles to be a kaon and the others pions.
In addition the kaon momentum has to exceed 2 GeV/$c$ for the $D^+$ decay, the
leptonic modes and the decays with $\pi^0$ reconstruction,
and 1 GeV/$c$ for all the other decay channels. A $D^{\ast+}$ candidate
%LLLLLLLLLL   Do you mean 'GT 0.4' For a LOW momentum pion?
is obtained by associating a low momentum pion down to 0.4 GeV/$c$ 
to the reconstructed $D^0$ meson. The charge of the pion is required to be
opposite to that of the kaon from the $D^0$ decay.

For the semileptonic decay modes $D^0 \rightarrow K^-e^+\nu$ and
$D^0 \rightarrow K^-\mu^+\nu$, the lepton is required to be identified,
using standard DELPHI identification criteria \cite{performance,leptid}. 

For the reconstruction of $\pi^0 \rightarrow \gamma\gamma$ decays,
three different classes of candidates measured in the HPC are used \cite{performance,pi0}. 
The measurement of two separated photon showers in the HPC are used to 
construct the
$\pi^0$. Above $\pi^0$ energies of 6--8\,GeV, the angle between the
$\gamma\gamma$ pair is too small to
separate the photon showers. The pion is therefore derived from the analysis 
of the shower shape of these merged photons. Information from
photons converted in front of the TPC is also used to reconstruct photons and 
thereby the neutral pion.

The particle identification provided by the RICH and the
specific energy loss $dE/dx$ measurement in the TPC are used to 
reduce the combinatorial background.
Due to the large number of pions in the hadronic final state,
combinations in which a pion is assigned as a kaon candidate are the
main contribution to the background. 
To optimize the efficiency of the $D$ signal, a pion veto, rather than  
direct kaon identification, has been introduced.
Tagging is performed using DELPHI standard tagging routines for the
RICH~\cite{newtag} and the
$dE/dx$~\cite{performance} identification. For the RICH, the measured Cherenkov
angle information is translated into $\pi$, $K$ and $p$ tagging information
${\rm tag}_{\rm Rich}^{\pi,K,p}$, taking into account the quality of the 
measurement. The way the information from the two radiators is combined
depends on the momentum of the candidate, in order to guarantee the best separation
over almost the whole momentum range.
Kaon candidates are tagged if they have either no pion tag or a very loose
pion tag ${\rm tag}_{\rm Rich}^{\pi}$.

The $dE/dx$ information is 
used only if no RICH information is available.
For each track a probability $P$ can be expressed in terms of the expected
ionisation for a given particle hypothesis:
\begin{equation}
  P_{K,\pi} = exp \left\{  -\frac{1}{2}\left(\frac{dE/dx-dE/dx_{K,\pi}}{\sigma_{K,\pi}}
  \right)^2 \right\}.
\end{equation}
This can be translated into a normalized kaon probability:
\begin{equation}
{\rm tag}_{\rm TPC} = \frac{P_{K}}{P_{K}+P_{\pi}}.
\end{equation}
Depending on the decay channel,
a cut ${\rm tag}_{\rm TPC} > 0.2-0.3$ to the kaon candidate is applied.

For the $D^0$ and the $D^+$ decays the kaon has to be tagged by the RICH or the
TPC as explained above. For the $D^{\ast +}$ decay channels a tag of the kaon candidate is not necessarily
required because of the cut on the mass difference $\Delta {\rm m}={\rm
  m}_{D^{\ast +}}-{\rm m}_{D^{0}}$ which selects rather pure sample of $D$ mesons.

For all decay modes a secondary vertex fit for the $D^{0/+}$ is performed
and the $D^{0/+}$ flight distance and improved track parameters are
obtained. All tracks associated
to a $D$ are required to have at least one hit in the vertex detector.
A further reduction of background for the $D^+$ is achieved
by rejecting track combinations with a probability for $\chi^2$ of the vertex
fit less than
0.001. The slow pion from the $D^{\ast+}$
decay is constrained to the $D^0$ vertex, which is a good approximation for
the $D^{\ast+}$ decay vertex because of the small transverse
momentum of the slow pion with respect to the flight direction  of
the $D^0$. 

\begin{table}[hbt]
  \begin{center}
    \renewcommand{\arraystretch}{1.15}
    \begin{tabular}{|l|lcl|r@{.}l||c|r@{.}l||c|c|}
    \hline
%------------------------------------------------------------------------------
~~~~~decay mode& \multicolumn{3}{c|}{$\Delta L$ for $D$ [cm]}
&\multicolumn{2}{c||}{$X_E^{cut}$}& 
  $a$& \multicolumn{2}{c||}{$b$} &
   $c$ & $d$\\
  \hline
%------------------------------------------------------------------------------
$D^{\ast+} \rightarrow (K^-\pi^+)\pi^+$              
&-0.05 &to& 2.0 & 0&15& -2.5 & 0&04 & 3.0 & 0.1  \\
%-----
$D^{\ast+} \rightarrow (K^-\pi^+\pi^-\pi^+)\pi^+$    
&0.0 &to& 2.0   & 0&3 & -1.0 & 0&05 & 2.0 & 0.1\\
%-----
$D^{\ast+} \rightarrow (K^-\pi^+\gamma\gamma)\pi^+$  
&0.0 &to& 2.0   & 0&3 & -1.5 & 0&09 & 2.0 & 0.1\\
%-----
$D^{\ast+} \rightarrow (K^-\mu^+\nu)\pi^+$           
&0.0 &to& 2.0   & 0&2 & -1.2 & 0&06&  3.0 & 0.1  \\
%-----
$D^{\ast+} \rightarrow (K^-e^+\nu)\pi^+$             
&0.0 &to& 2.0   & 0&2 & -1.4 & 0&05&  3.0 & 0.1 \\
%-----
$D^{\ast+} \rightarrow (K^-\pi^+(\pi^0))\pi^+$       
&0.0 &to& 2.0   & 0&2 & -2.5 & 0&03&  3.0 & 0.0 \\
%-----
$D^{+}~ \rightarrow K^-\pi^+\pi^+$                   
&0.125 &to& 2.0 & 0&35& 
--- & \multicolumn{2}{c||}{---}&  3.0 & 0.1 \\
%-----
$D^0~~ \rightarrow K^-\pi^+$                         
&0.05 &to& 2.0  & 0&3 & -0.5 & 0&125&  2.0 & 0.2 \\
%-----
$D^0~~ \rightarrow K^-\pi^+(\pi^0)$                  
&0.05 &to& 2.0  & 0&3 & -0.6 & 0&15&  2.0 & 0.2 \\
%-----
  \hline 
%------------------------------------------------------------------------------
    \end{tabular}
  \end{center}
  \caption{\label{xedlcuts} The selection cuts for $X_E$ versus $\Delta L$
    ($a,b$), $X_E$ versus $\cos\theta_H$ ($c,d$) and the absolute cut on
    $X_E$.}
\end{table}

Cuts on the helicity angle distribution are used
to achieve a further  significant reduction of the combinatorial background.
The helicity angle $\theta_H$ is defined as the angle of the
sphericity axis in the $D^{0/+} $ rest frame with respect to the $D^{0/+} $
flight direction. The orientation of the sphericity axis corresponds to the
kaon candidate in the decay. $D^{0/+} $ decays are isotropic in
$\cos\theta_H$, whereas the background is 
%extremely 
peaked at
$\cos\theta_H = \pm 1$. Because of the shape of the energy spectrum of charged particles in
hadronic $Z$ events, the combinatorial background is concentrated at small
scaled energies $X_E(D)$. Therefore $X_E$  cuts which depend on the helicity 
angle are used, in order to remove higher background
contributions at small $D$ meson energies: 
\begin{equation}
X_E > 0.5 \cdot  e^{c  (|cos\theta_H| - 1)} + d\,.
\end{equation}

\noindent
In addition the scaled energy $X_E$ of the $D$ combination is required
to exceed the limits $X_E^{cut}$ given in table \ref{xedlcuts}, where
the parameters $c$ and $d$ are also shown.

The distance between the primary and the $D^{0/+} $ vertex is calculated in
the plane transverse to the beam axis and projected onto the $D^{0/+} $ direction of flight to
obtain the decay length $\Delta L$. A vertex combination
is accepted if $\Delta L$ is within the range specified in table \ref{xedlcuts}
for the different decay modes. In addition an $X_E$ dependent cut on $\Delta
L$ is applied:
%In addition a cut is applied in the plane
%of $X_E$ and $\Delta L$ to 
\begin{equation}
\Delta L(X_E) > a \cdot (X_E-X_E^{cut})^2 + b\,.
\end{equation}

\noindent
This rejects combinatorial background, which is 
concentrated at low values of both of these variables.
The parameters $a$ and $b$ are given in table \ref{xedlcuts}.

\begin{table}[hbt]
    \renewcommand{\arraystretch}{1.15}
  \begin{center}
    \begin{tabular}{|l|c|c|}
    \hline
%------------------------------------------------------------------------------
%%%CD   &  &  \\
 mode & $D^{0/+}$ mass interval & max.$\Delta$m  \\ 
%-----
  & [GeV/$c^2$] & [GeV/$c^2$]  \\ \hline
%------------------------------------------------------------------------------
$D^{\ast+} \rightarrow (K^-\pi^+)\pi^+$ &
 1.790 to 1.94 & 0.160  \\
%-----
$D^{\ast+} \rightarrow (K^-\pi^+\pi^-\pi^+)\pi^+$  &
 1.845 to 1.90 & 0.160  \\
%-----
$D^{\ast+} \rightarrow (K^-\pi^+\gamma\gamma)\pi^+$ &
 1.740 to 1.98 & 0.165 \\
%-----
$D^{\ast+} \rightarrow (K^-\mu^+\nu)\pi^+$  &
 0.750 to 1.75 & 0.250  \\
%-----
$D^{\ast+} \rightarrow (K^-e^+\nu)\pi^+$ &
 0.750 to 1.75 & 0.250  \\
%-----
$D^{\ast+} \rightarrow (K^-\pi^+(\pi^0))\pi^+$  &
 1.350 to 1.75 & 0.175  \\
%-----
$D^{+}~ \rightarrow K^-\pi^+\pi^+$  &
 1.700 to 2.05 & -  \\
%-----
$D^0~~ \rightarrow K^-\pi^+$  &
 1.750 to 2.20 & -  \\
%-----
$D^0~~ \rightarrow K^-\pi^+(\pi^0)$  &
 1.500 to 1.70 & -  \\
%-----
  \hline 
%------------------------------------------------------------------------------
    \end{tabular}
  \end{center}
  \caption{\label{masscuts} Mass and mass difference cuts for the selection
of the $D$ meson signal plus sideband regions.}
\end{table}

For the $D^{+} \rightarrow K^-\pi^+\pi^+$ mode, a 
cut\footnote{
The mass difference for the $D^{+}$ is given as 
$\Delta\mbox{m}=\mbox{m}_{D^{+}}-\mbox{m}_{K^-\pi^+}$, in analogy with that for
the $D^{\ast}$ decay: $\Delta\mbox{m}=\mbox{m}_{D^{\ast}}-\mbox{m}_{D^{0}}$.
Both possible $\mbox{m}_{K^-\pi^+}$ combinations are tested. } of
$\Delta$m $> 200$ Mev
is used to veto $D^{\ast+}$ decays. 

For the $D^{\ast+} \rightarrow (K^-\pi^+\pi^0)\pi^+$ decay mode with a
reconstructed $\pi^0$, an additional cut on the $D^0$  Dalitz plot  is applied 
to select the dominant decay via $D^0 \to K^-\rho^+$:\\
\begin{center}
\begin{tabular}{ccc}
0.5~GeV/$c^2$ $<$ $\mbox{m}_{K^-\pi^+}$ $<$ 1.1~GeV/$c^2$ & ~~and~~ & 1.4~GeV/$c^2$ $<$
  $\mbox{m}_{K^-\pi^0}$ $<$ 1.8~GeV/$c^2$ \\ 
1.4~GeV/$c^2$ $<$ $\mbox{m}_{K^-\pi^+}$ $<$ 1.8~GeV/$c^2$ & ~~and~~ & 0.5~GeV/$c^2$ $<$
  $\mbox{m}_{K^-\pi^0}$ $<$ 1.1~GeV/$c^2$ \\ 
\end{tabular} \\
\end{center}
%The Dalitz plot including the cuts is shown in figure
%\ref{fig_dalitz}. To illustrate the real $D^{\ast+}$ decays the prediction
%from the simulation is superimposed.

\begin{figure}
 \begin{minipage}[t]{7.8cm}
  \mbox{\epsfig{file=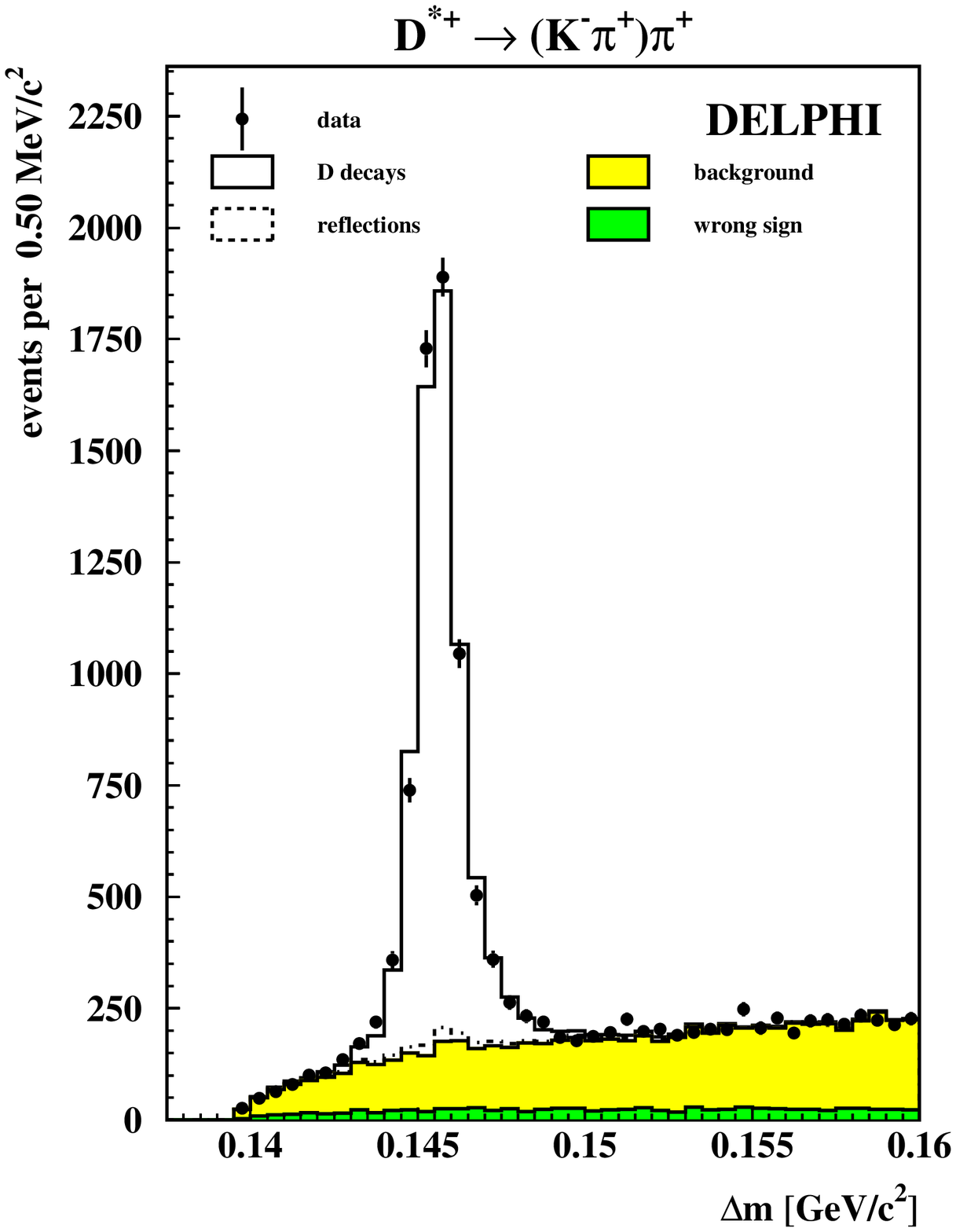,width=8.2cm}}
 \end{minipage} \hfill
 \begin{minipage}[t]{7.8cm}
  \mbox{\epsfig{file=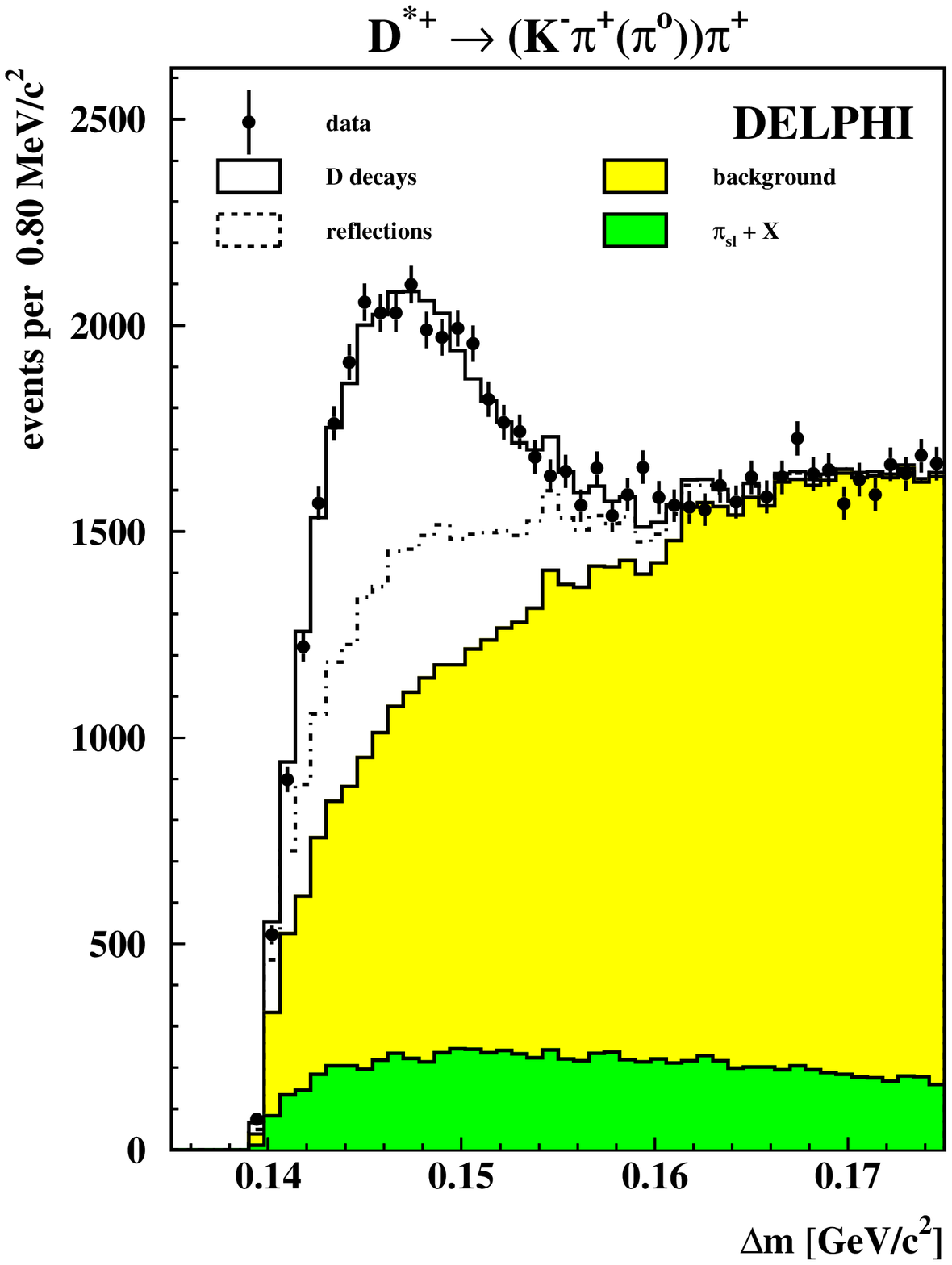,width=8.2cm}}
 \end{minipage}\\
 \begin{minipage}[t]{7.8cm}
  \mbox{\epsfig{file=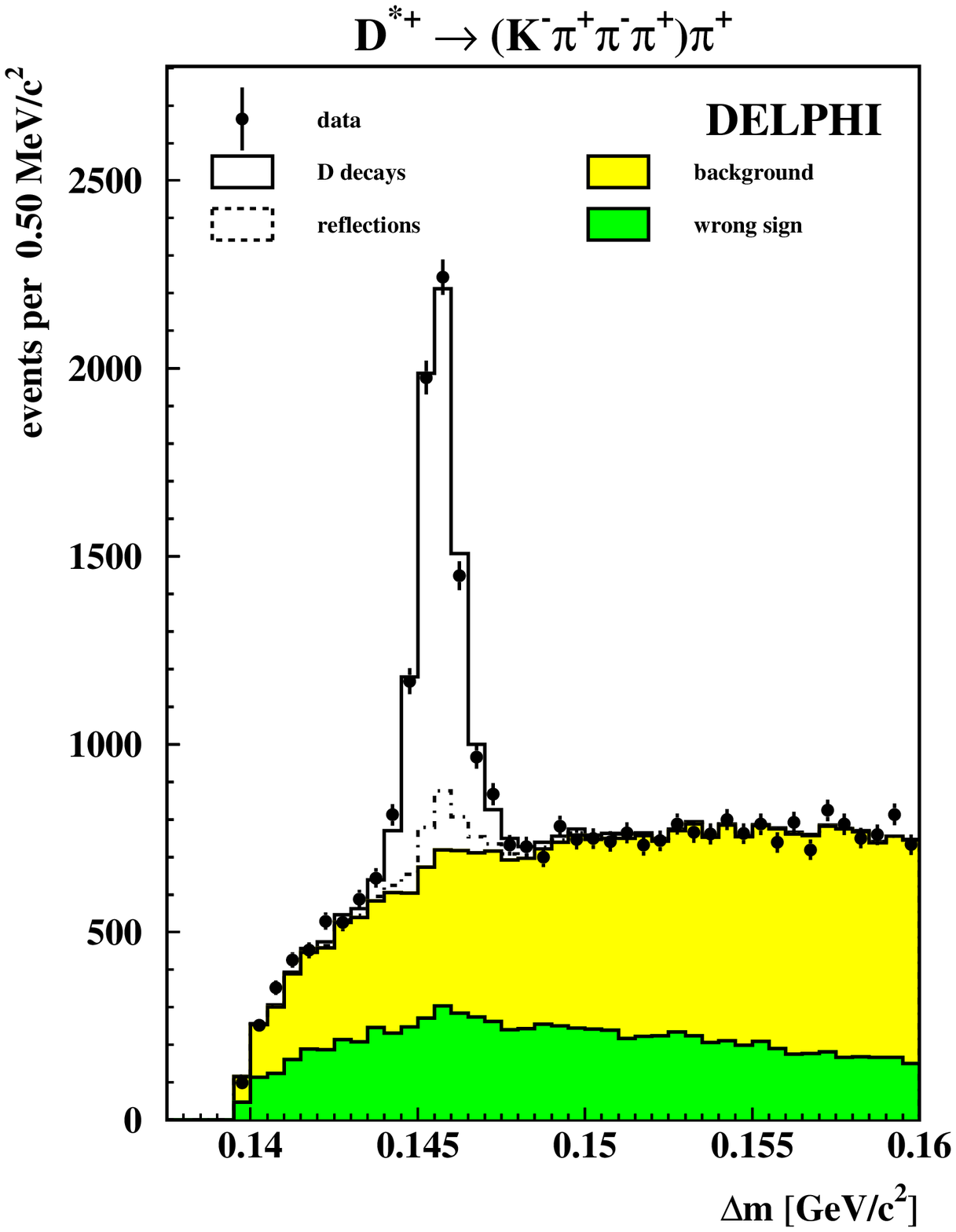,width=8.2cm}}
 \end{minipage} \hfill
 \begin{minipage}[t]{7.8cm}
  \mbox{\epsfig{file=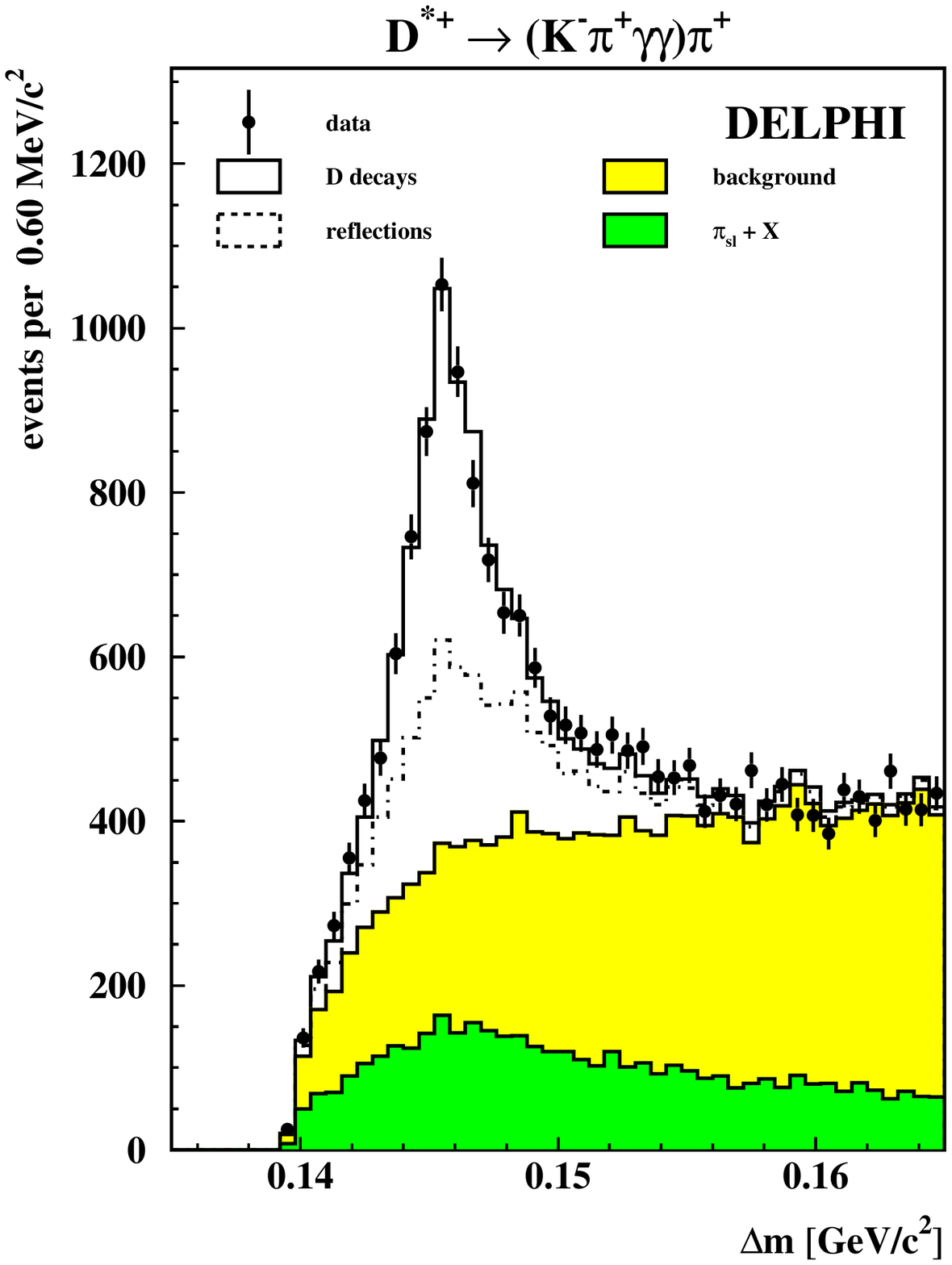,width=8.2cm}}
 \end{minipage}
 \caption[]{\label{fig_dm1} The mass difference distributions $\Delta$m
    for the different decay modes. $\Delta$m is defined as the
    difference between the mass of the $D^{\ast+}$ and the $D^0$ candidate.
    The data are compared to the simulation. Contributions from
    reflections, partially reconstructed $D^{\ast+}$ decays ($\pi_{sl} + X$)
    and combinatorial background are also shown. See section \ref{measurement}
    for the discussion of these contributions.}
\end{figure}

\begin{figure}
  \begin{minipage}[t]{7.8cm}
    \mbox{\epsfig{file=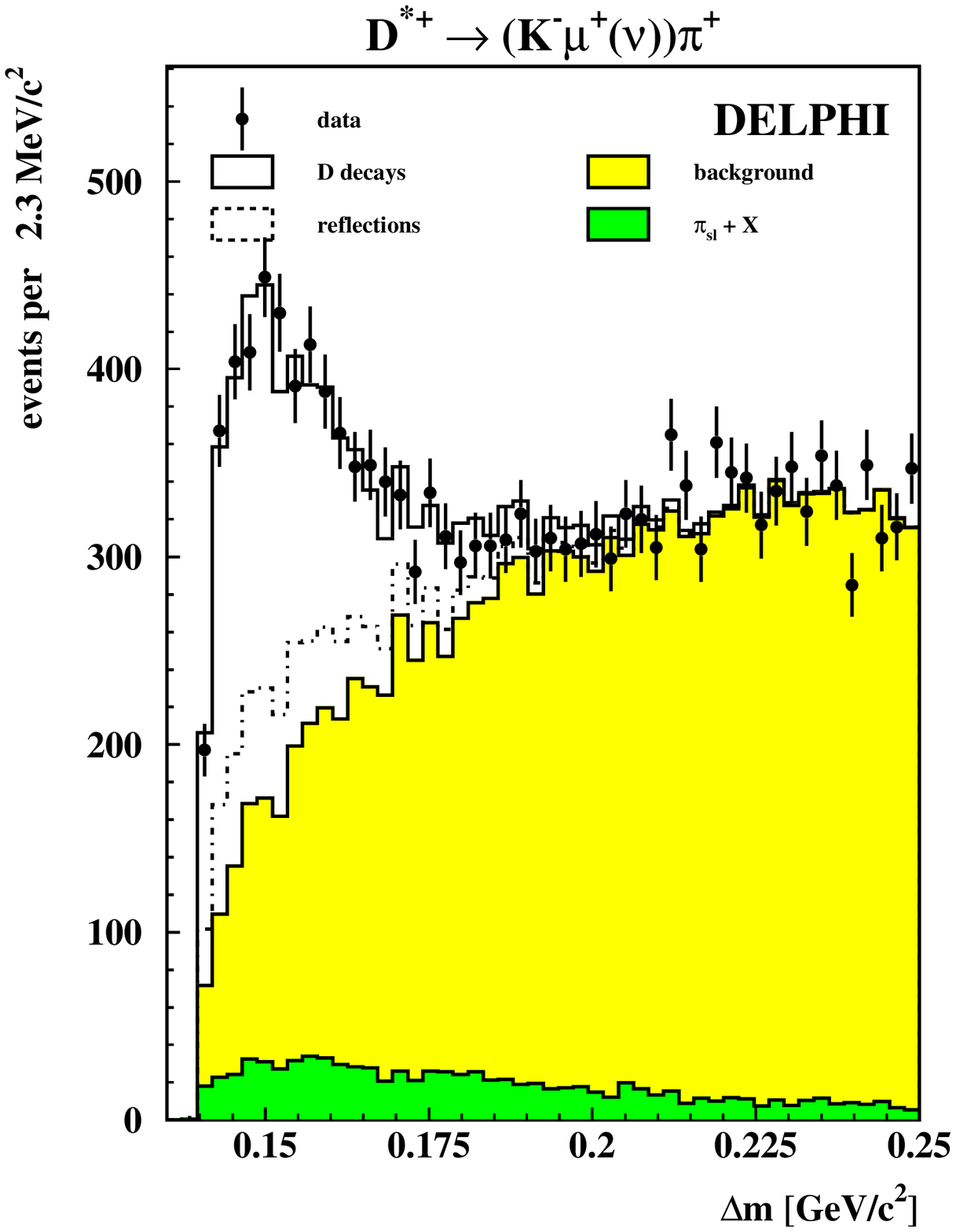,width=8.2cm}}
  \end{minipage} \hfill
  \begin{minipage}[t]{7.8cm}
    \mbox{\epsfig{file=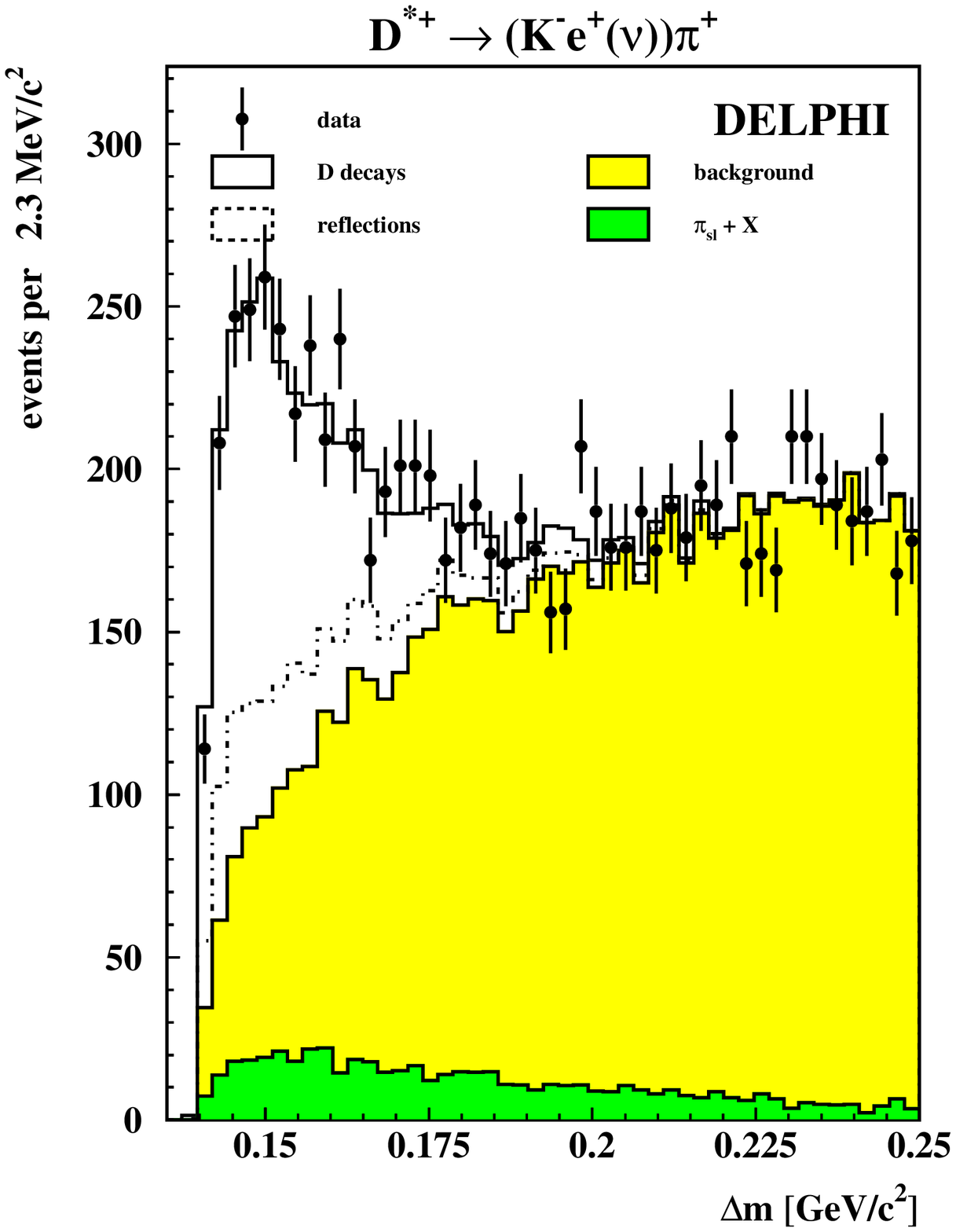,width=8.2cm}}
  \end{minipage}
  \begin{minipage}[t]{7.8cm}
    \mbox{\epsfig{file=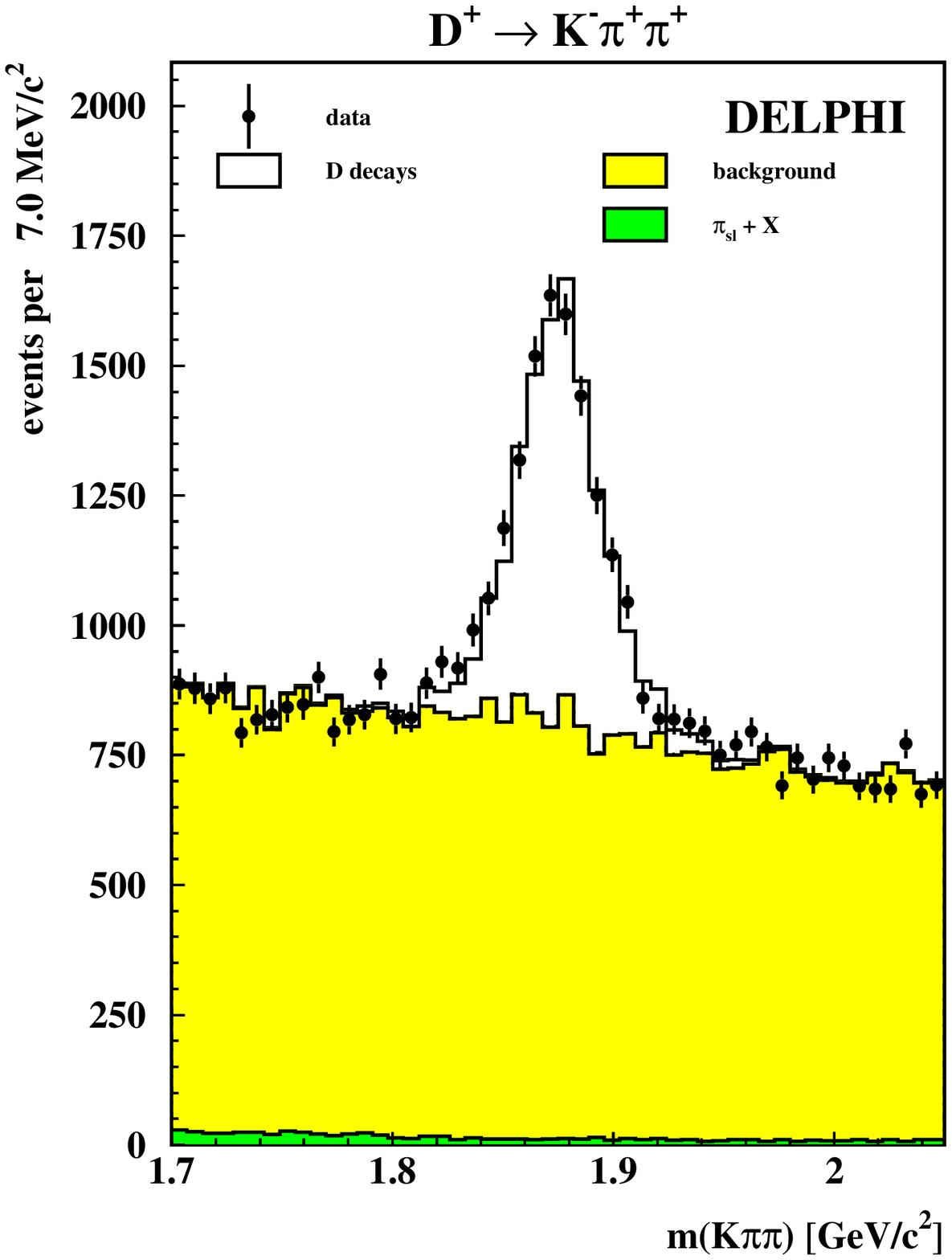,width=8.2cm}}
  \end{minipage} \hfill
  \begin{minipage}[t]{7.8cm}
    \mbox{\epsfig{file=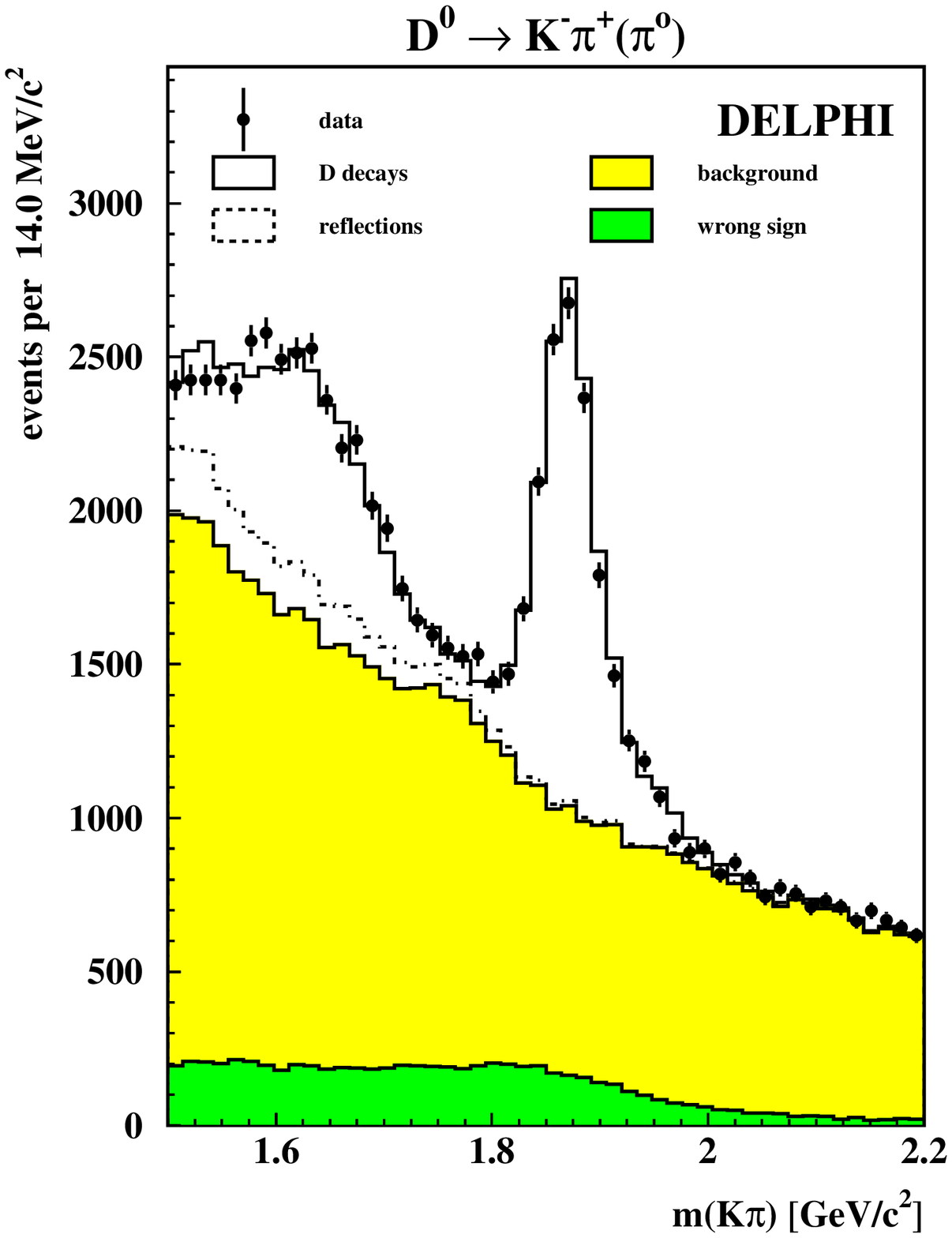,width=8.2cm}}
  \end{minipage}
 \caption[]{\label{fig_dm2} The mass difference distributions $\Delta$m
    for the semileptonic decay modes are shown in the two upper plots. $\Delta$m is defined 
    as the difference between the mass of the $D^{\ast+}$ and
    the $D^0$ candidate.
    The data are compared to the simulation. Contributions from
    reflections, partially reconstructed $D^{\ast+}$ decays and combinatorial
    background are also shown.
    The lower diagrams show the $D^+$ and $D^0$ mass 
    distributions. For the $D^0$ the background distribution for
    candidates with wrong mass assignments is also shown. In the $D^0$ case, 
    the peak in the $K^-\pi^+$ mass distribution comes from $D^0$ decays into 
    $K^-\pi^+$, and the broad enhancement at lower values is from the
    $K^-\pi^+\pi^0$ decay.}
\end{figure}

The mass bands to select the different $D^{0/+} $ decay modes
and the cuts on the mass difference are listed in table \ref{masscuts}. 
The  mass and mass difference bands in the signal 
regions for the different decay channels used in the final analysis 
are given in table \ref{samples}; the sidebands are defined as the mass or
mass difference regions given in table \ref{masscuts} not selected by the cuts
in table \ref{samples}. The
mass differences for the $D^{\ast+}$ and also the $D^0$ and $D^+$ mass distributions 
are shown in figures \ref{fig_dm1} and \ref{fig_dm2}.
The wrong sign combination for
$D^0 \rightarrow K^-\pi^+$, where the kaon and pion masses are interchanged,
%where the assignment of the kaon mass is used
%for the pion, 
is also shown in figure \ref{fig_dm2}.
The histograms show the simulated distributions normalized to the data samples.
The contributions of signal and background are adjusted to compensate for
different $D$ rates in data and simulation.

%=========================================================================%
\section{\label{measurement} Measurement of {\boldmath $A_{FB}^c$}
and {\boldmath $A_{FB}^b$}}
%=========================================================================%

For a measurement of \Afbc\ and \Afbb\ from the polar angle
$\cos\theta_{thrust}$ of the thrust axis in the $D$ meson events, it is
necessary to separate $D$ from $c\bar{c}$ and $b\bar{b}$ events and the
combinatorial background. Since the $c$ and $b$ asymmetries are expected
to be of comparable size and to have the same relative sign, the statistical
precision of the measurement is limited by the
negative correlation between both asymmetries. In this analysis,  good
separation with a small correlation is obtained by using
the scaled energy distribution $X_E$ of the $D$ candidates and the
event $b$-tagging variable ${\cal P}_{ev}$ \cite{delphi_rb}. %\cite{combtag}.

The hadronization of primary $c$ quarks leads to high energy
$D$ mesons, whereas $b$ quarks fragment into
$b$--hadrons which then decay into $D$ mesons with a softer energy
spectrum. The combinatorial background is concentrated at
low $X_E$. Furthermore $b\bar b$ events can be identified by $b$--tagging,
which utilises special features of $B$ hadrons, as compared with other hadrons.
%, for example  
%by the long lifetime of $B$ hadrons, compared to the shorter lifetime of $D$ 
%mesons.
%
%The $b$-tag takes advantage of specialties of $B$ hadrons w.r.t. all other
%particles. 
The combined $b$-tagging
%\cite{combtag}
used in this analysis takes into account the long lifetime
and the large mass of $B$ hadrons,
their higher decay multiplicity and their large $X_E(B)$.
This leads to a high tagging efficiency in combination with
good separation power of the tag.
%\begin{figure}
%\begin{center}
%%  \begin{minipage}[t]{7.8cm}
%    \mbox{\epsfig{file=dalitz_94_c.eps,width=13.cm}}
%%  \end{minipage} \hfill
% \caption[]{\label{fig_dalitz} Dalitz plot for the decay $\ds \to \dnpi \to
%   \kpipinpis$. The framed areas correspond to the cut values. To demonstrate
%   the signal regions, the prediction for the decay from the simulation is
%   superimposed in black.}
%\end{center}
%\end{figure}

The shape of the combinatorial background is tested using
the sidebands in the mass (or the mass difference) distribution. Due to
the different relative acceptance of $D$ mesons and background at small and
large polar angles, the fit method has to take into account  the
$|\cos\theta_{thrust} |$ dependence of the different classes.

The charge state of a signal $D$ is directly correlated to the charge 
of the primary quark, whereas the charge correlation of the combinatorial
background is expected to be very small. 

For the asymmetry measurement, partially reconstructed $D^{\ast+}$ mesons
($\pi_{sl} + X$) and reflections from other decay modes (see figures
\ref{fig_dm1} and \ref{fig_dm2}) have to be considered as signal to avoid
charge correlations in the background. The contributions from reflections,
where some particles
from the $D$ decay are assigned a wrong mass or are missing, and true
$D$ decays are treated as one class, because of 
the similar shape of the signals and the charge correlation with the primary
quark. This leads to a significant increase of the sample for the
$K^-\pi^+(\pi^0)$ decay mode.
The rate of partially reconstructed $D^{\ast+}$ mesons, where a $\pi^+$ from a
$D^{\ast+}$ decay is combined with a fake $D^0$,
depends on the branching ratio $D^{\ast+} \rightarrow D^0\pi^+$, the
$D^{\ast+}$ production rate and the efficiency in the relevant mass difference
interval. The contribution of partially reconstructed $D^{\ast+}$ decays to
the signal is taken from the simulation and contributes to the systematic
uncertainty. In the case of  $D^0 \rightarrow K^-\pi^+ / K^-\pi^+(\pi^0)$
decay modes without the $D^{\ast+}$ constraint, candidates with wrong mass
assignments flip the sign of the estimated primary quark direction. This is
taken into account in the fit; the systematic error allows for uncertainties.

To avoid double counting of events, only one $D$ candidate in the signal region
per event is retained. For a given event the $D$ candidate with the largest
kaon momentum is used. If two candidates for a given decay mode use
the same kaon track, the one with the largest $X_E(D)$ is used.
Events entering the signal region for the $K^-\pi^+$ decay
mode are removed from the $K^-\pi^+\pi^-\pi^+$ distribution and events
from both decay modes are then removed from the $K^-\mu^+\nu$ or
$K^-e^+\nu$ distribution and so forth. The order of the rejection is as listed
in table \ref{masscuts}. Good agreement between data and simulation was 
found for the rejection.

The numbers of reconstructed $D$ decays given in table \ref{samples}
are  obtained from fits to the mass spectra. A total sample of $61829 \pm 521$
reconstructed $D$ decays is used for the asymmetry measurement.
The $D$ mass bands to select $D$ meson candidates are listed
in table \ref{samples}.

\begin{table}[htb]
    \renewcommand{\arraystretch}{1.15}
  \begin{center}
    \begin{tabular}{|l|r@{~$\pm$~}r|c|c|c|}
\hline
     & \multicolumn{2}{c|}{}  & signal region & signal region &
 \\
~~~~decay mode &    \multicolumn{2}{c|}{signal events}     
& $\Delta$m [GeV/$c^2$] & m$_{D}$ [GeV/$c^2$] &
$R_{S/B}$ \\
\hline
  $D^{\ast+} \rightarrow (K^-\pi^+)\pi^+$             & 6030 &103&
  0.143-0.148   & -  & 0.95 $\pm$ 0.02 \\
  $D^{\ast+} \rightarrow (K^-\pi^+\pi^-\pi^+)\pi^+$   & 5123 &103 &      
  0.143-0.148   & -  & 0.86 $\pm$ 0.02 \\
  $D^{\ast+} \rightarrow (K^-\pi^+\gamma\gamma)\pi^+$ & 5787 &125&
  0.141-0.151   & -  & 1.19 $\pm$ 0.03 \\
  $D^{\ast+} \rightarrow (K^-\mu^+\nu)\pi^+$          & 3042 &91&
  $< 0.180$     & -  & 0.64 $\pm$ 0.02 \\
  $D^{\ast+} \rightarrow (K^-e^+\nu)\pi^+$            & 1810 &65&
  $< 0.180$     & -  & 0.98 $\pm$ 0.04 \\
  $D^{\ast+} \rightarrow (K^-\pi^+(\pi^0))\pi^+$      & 15111&232&
  $< 0.152$     & -  & 1.16 $\pm$ 0.02 \\
  $D^{+}~ \rightarrow K^-\pi^+\pi^+$                  & 5667 &161&
   -            & 1.83-1.91 & 0.83 $\pm$ 0.02 \\
  $D^0~~ \rightarrow K^-\pi^+$                        & 9311 &232&
   -            & 1.80-1.93 & 1.00 $\pm$ 0.02 \\
  $D^0~~ \rightarrow K^-\pi^+(\pi^0)$                 & 9948 &298&
   -            & 1.50-1.70 & 1.21 $\pm$ 0.04 \\
%-----------------------------------------------------------------------
\hline
    \end{tabular}
  \end{center}
  \caption{\label{samples} $D$ meson samples used for the measurement,
           cuts to select signal regions, and the relative normalizations
           $R_{S/B}$ of signal to background for data and simulation.}
\end{table}

%=========================================================================%
\subsection{\label{chi2} The minimum {\boldmath $\chi^2$} fit}
%=========================================================================%

\begin{figure}[p]
  \begin{minipage}[t]{7.8cm}
    \mbox{\epsfig{file=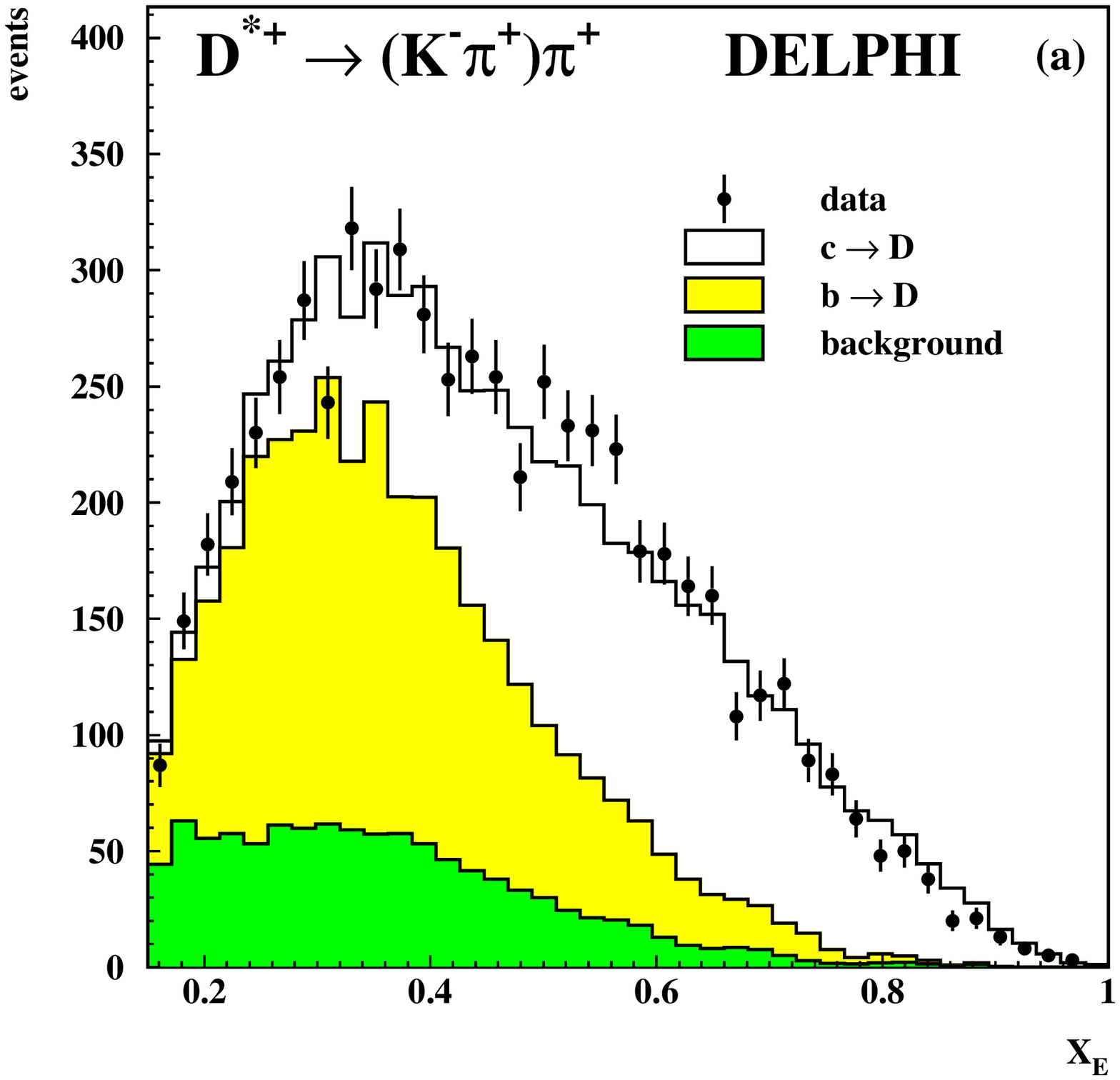,width=8.5cm}}
  \end{minipage} \hfill
  \begin{minipage}[t]{7.8cm}
    \mbox{\epsfig{file=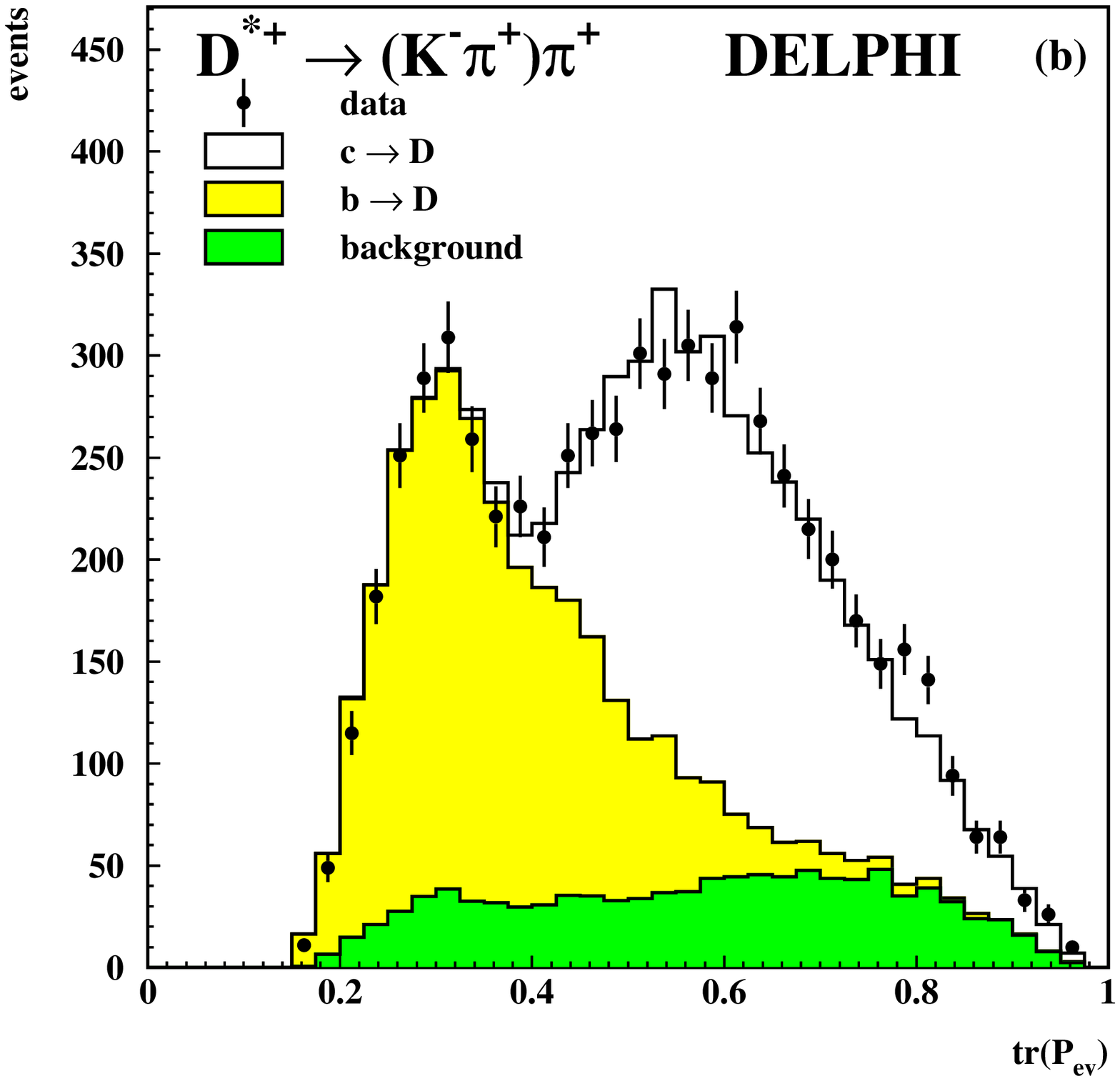,width=8.5cm}}
  \end{minipage}
  \begin{minipage}[t]{7.8cm}
    \mbox{\epsfig{file=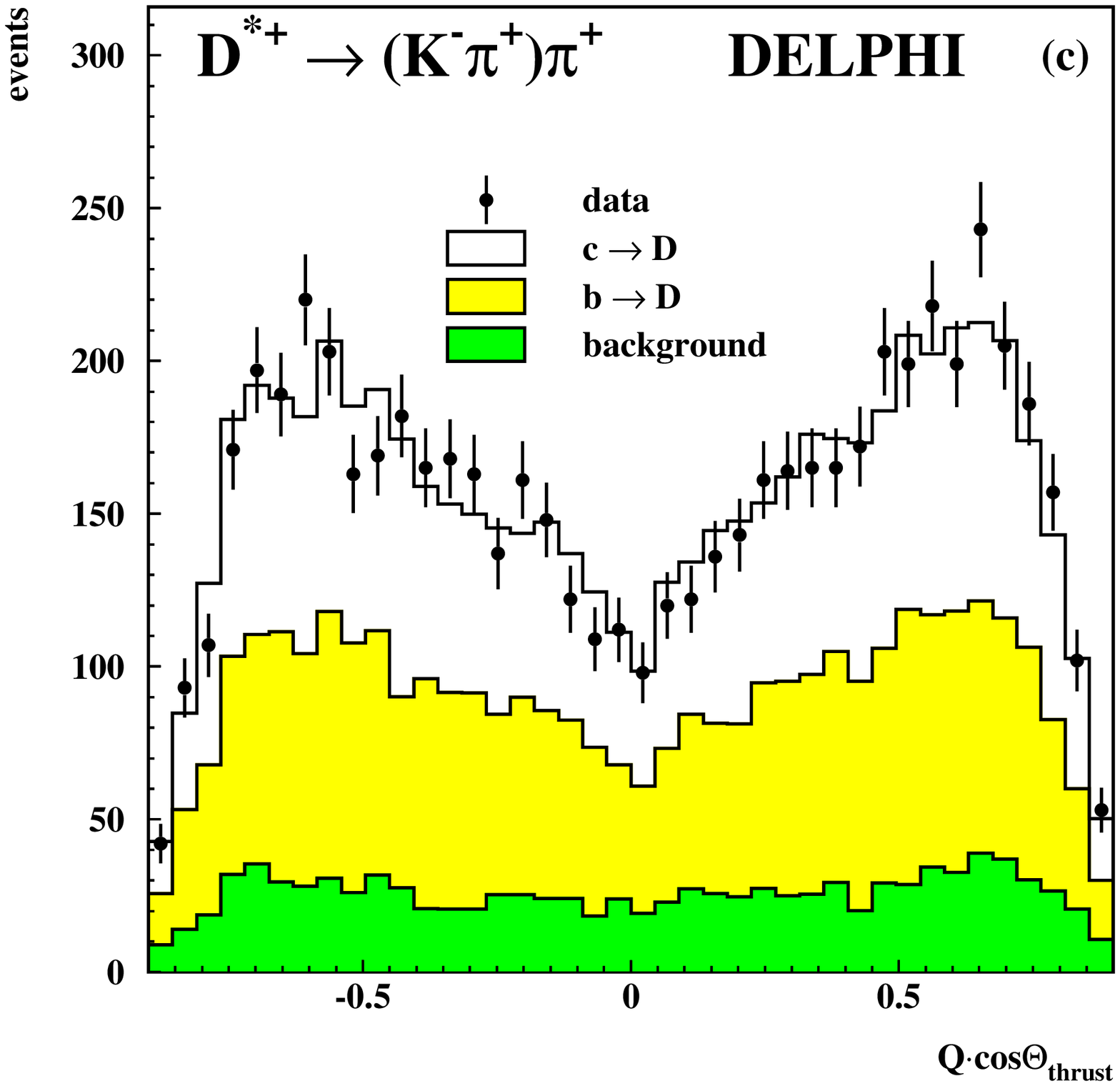,width=8.5cm}}
  \end{minipage} \hfill
  \begin{minipage}[t]{7.8cm}
  \end{minipage}
\caption[]{\label{fit_kpi} The $X_E$, $tr({\cal P}_{ev})$ and
  $Q\cdot\cos(\theta_{thrust})$  
         distribution for the signal region of the $D^{\ast+} \rightarrow
         (K^-\pi^+)\pi^+$ decay mode. The data are compared to the simulation
         where $D^{\ast+}$ from charm and bottom events and combinatorial
         background are shown separately. See section \ref{chi2} for the
         definition of $tr({\cal P}_{ev})$.}
\end{figure}

The determination of the asymmetries at $\sqrt{s} = 91.235$\,GeV is achieved by a minimum $\chi^2$
fit to the $D$ samples using the scaled energy $X_E$, the transformed $b$-tagging
variable  $tr({\cal P}_{ev})$ for the event 
%\cite{combtag}
and the polar angle
$Q\cdot\cos\theta_{thrust}$ signed with the charge state $Q$ of the $D$.
Examples of these distributions for the 
$D^{\ast+} \rightarrow (K^-\pi^+)\pi^+$ channel are shown in figure
\ref{fit_kpi}. The measured distributions are compared to the predictions of
the simulation, split into charm, bottom and background events. The simulated
prediction is normalized to the data to reproduce the signal to background
ratio. Therefore a factor $R_{S/B}$ (see table \ref{samples}) is introduced for
each decay mode, which compensates for different $D$ rates in data
and simulation. After this correction, good agreement is found in all
distributions. The shape of the background distribution, as obtained from
the sidebands, is well reproduced by the simulation.

\begin{table}[t]
    \renewcommand{\arraystretch}{1.15}
  \begin{center}
    \begin{tabular}{|l|c|c|c|c|}
      \hline
Decay mode & \multicolumn{3}{|c|}{number of bins per } &
                                                         average number \\
$\sqrt{s} = 91.235$\,GeV    & $X_E$ & $tr({\cal P}_{ev})$ & $|\cos\theta_{thrust}|$ & of events \\
      \hline
$D^{\ast+} \rightarrow (K^-\pi^+)\pi^+$             &  4  &  5  &  4  & 77.8\\
$D^{\ast+} \rightarrow (K^-\pi^+\pi^-\pi^+)\pi^+$   &  5  &  5  &  5  & 65.8\\
$D^{\ast+} \rightarrow (K^-\pi^+\gamma\gamma)\pi^+$ &  4  &  4  &  4  & 88.0\\
$D^{\ast+} \rightarrow (K^-\mu^+\nu)\pi^+$          &  4  &  6  &  4  & 49.6\\
$D^{\ast+} \rightarrow (K^-e^+\nu)\pi^+$            &  4  &  4  &  3  & 56.4\\
$D^{\ast+} \rightarrow (K^-\pi^+(\pi^0))\pi^+$      &  6  &  7  &  5  & 77.7\\
$D^{+}~ \rightarrow K^-\pi^+\pi^+$                  &  5  &  5  &  5  & 87.9\\
$D^0~~ \rightarrow K^-\pi^+$                        &  5  &  6  &  5  & 80.0\\
$D^0~~ \rightarrow K^-\pi^+(\pi^0)$                 &  7  &  7  &  7  & 59.9\\
      \hline
    \end{tabular}
  \end{center}
  \caption[]{Number of bins in each dimension used for the individual
           decay modes and the average number of data events per bin
           at $\sqrt{s} = 91.235$\,GeV.}
  \label{binning}
\end{table}

A transformation of the event variable $P_{ev}$ is used for the $b$-tagging
distribution:

\begin{eqnarray}
tr({\cal P}_{ev}) = \frac{2.5}{5.1+{\cal P}_{ev}}.
\end{eqnarray}

\noindent
The bins in the three dimensional
$X_E$, $tr({\cal P}_{ev})$ and $\cos\theta_{thrust}$ space have
been chosen such that each bin contains about 70 events (table \ref{binning}).
In each bin $i$ the differential asymmetry:

\begin{equation}
A^{obs,i}_{FB}=\frac{N_i^{+}-N_i^{-}}{N_i^{+}+N_i^{-}}
\end{equation}

\noindent
is calculated from the numbers of events $N_i^{+}$ and $N_i^{-}$ with
$Q\cdot\cos\theta_{thrust}$ greater or less than zero, respectively. The
observed asymmetry receives
contributions from $c$, $b$ and combinatorial background. The fractions
$f_{ji}$ of $D$ signal and reflection from $c$ and $b$ events as well
as the fractions of partially
reconstructed $D$ mesons and combinatorial background are taken from the 
simulation. Furthermore the combinatorial background is divided into $c$, 
$b$ and $uds$ contributions to account for the small charge correlation to the
primary quark in the
background, especially for the semileptonic decay mode of the $D^0$. This leads
to three different contributions 
%LLLLLLLLLLLLLLLLLLL
(i.e. signal plus reflections, partially reconstructed $D$, and 
combinatorial background)
from each of $c$ and $b$, and one for the
background from $uds$. 

The $\chi^2$ to be minimized is given by: 

\begin{equation}
\label{chi2-eqn}
\chi^2=\sum\limits_{i=1}^{N_{bins}}
       \left\{ A^{obs,i}_{FB} -
              \sum\limits_{j=1}^7 f_{ji}\,C_{ji}\,A_{FB}^j (\cos\theta_i)
              \right\}^2 / \sigma_i^2
\end{equation}

\noindent
where $\sigma_i$ accounts for the statistical error of both data and simulation;
$A_{FB}^j(\cos \theta_i)$ is the differential asymmetry:

\begin{equation}
A_{FB}^j(\cos \theta_i) = \frac{8}{3} \, A_{FB}^j \,
                   \frac{\cos\theta_i}{1+\cos^2\theta_i} ,
\end{equation}

\noindent
of $b$, $c$ or $uds$ events; 
%In each bin the average polar angle
%$\cos\theta_f$ of the primary quarks after photon radiation from the simulation
%is used. The emission of hard gluons in the following parton shower can change
%significantly the polar angle distribution of the primary quarks and thus the
%asymmetry information. The use of $\cos\theta_f$ takes this QCD corrections (see chapter \ref{sec-qcd}) into account
%in the fit. 
and $C_{ji}$ is the charge correlation of the class
$j$ calculated in each bin using the simulation. 
For $b$ events the mixing
effect leads to values of the charge correlation $C_{ji}$ which are smaller
than $1$, and thus to a smaller observed $b$
asymmetry. 
The simulation is used to estimate the mixing effect as
a function of the $b$--tagging $tr(P_{ev})$, because the $b$--tagging depends
on the individual $B$ lifetime
%%The simulation is used to estimate the effect of mixing which, because of
%%the $b$--tagging, depends on how long an individual $B$ lives
%LLLLLLLLLLL  I took out next sentence because it says the same as previous one
%The dependence of the mixing from the lifetime of the $B$ hadron 
%and thus from the 
%$b$-tagging is taken from the simulation
(see section \ref{mixing} for details).
The combinatorial background from $c$
is expected to have only a small charge correlation $C_{ji}$
to the primary quark at large energies $X_E$. The asymmetry of the
combinatorial background and of $D$ mesons from gluon splitting in $uds$ events
is expected to be very small. The predictions from the simulation for this
class in each
bin are subtracted in the fit. The two fit parameters $A_{FB}^c$ and 
$A_{FB}^b$ are used for
all three classes from $c$ and $b$, while for the single class from $uds$ the
prediction of the simulation is used. 
The agreement of data and simulation is tested
in the sidebands of the different samples where no significant deviations are
found.
\begin{table}[h]
    \renewcommand{\arraystretch}{1.15}
  \begin{center}
    \begin{tabular}{|l|c|c|c|r@{~~~~~~}|}
      \hline
Decay mode & \multicolumn{3}{|c|}{number of bins per } &
                                      \multicolumn{1}{|c|}{number of} \\
$\sqrt{s} = 89.434$\,GeV    & $X_E$ & $tr({\cal P}_{ev})$ & $|\cos\theta_{thrust}|$ & \multicolumn{1}{|c|}{candidates}  \\
      \hline
$D^{\ast+} \rightarrow (K^-\pi^+)\pi^+$             
&  3  &  3  &  4  & 378 \\%& 6387 \\
$D^{\ast+} \rightarrow (K^-\pi^+\pi^-\pi^+)\pi^+$   
&  4  &  4  &  3  & 539 \\%& 8705 \\
$D^{\ast+} \rightarrow (K^-\pi^+\gamma\gamma)\pi^+$ 
&  3  &  3  &  4  & 356 \\%& 5695 \\
$D^{\ast+} \rightarrow (K^-\mu^+\nu)\pi^+$          
&  3  &  3  &  4  & 260 \\%& 4576 \\
$D^{\ast+} \rightarrow (K^-e^+\nu)\pi^+$            
&  3  &  3  &  2  & 188 \\%& 3684 \\
$D^{\ast+} \rightarrow (K^-\pi^+(\pi^0))\pi^+$      
&  5  &  5  &  4  &1096 \\%&15135 \\
$D^{+}~ \rightarrow K^-\pi^+\pi^+$                  
&  4  &  4  &  4  & 814 \\%&12086 \\
$D^0~~ \rightarrow K^-\pi^+$                        
&  4  &  5  &  3  & 721 \\%&12811 \\
$D^0~~ \rightarrow K^-\pi^+(\pi^0)$                 
&  4  &  5  &  5  &1287 \\%&21407 \\
      \hline
    \end{tabular}
  \end{center}
  \caption[]{Number of bins in each dimension used for the individual
           decay modes and the  number of data events per bin
           at $\sqrt{s} = 89.434$\,GeV.}
  \label{pbinning}
\end{table}
\begin{table}[h]
    \renewcommand{\arraystretch}{1.15}
  \begin{center}
    \begin{tabular}{|l|c|c|c|r@{~~~~~~}|}
      \hline
Decay mode & \multicolumn{3}{|c|}{number of bins per } &
                                      \multicolumn{1}{|c|}{number of}  \\
$\sqrt{s} = 92.990$\,GeV    & $X_E$ & $tr({\cal P}_{ev})$ & $|\cos\theta_{thrust}|$ & \multicolumn{1}{|c|}{candidates} \\
      \hline
$D^{\ast+} \rightarrow (K^-\pi^+)\pi^+$             
&  3  &  4  &  4  & 576 \\%& 6387 \\
$D^{\ast+} \rightarrow (K^-\pi^+\pi^-\pi^+)\pi^+$   
&  4  &  4  &  4  & 734 \\%& 8705 \\
$D^{\ast+} \rightarrow (K^-\pi^+\gamma\gamma)\pi^+$ 
&  3  &  4  &  3  & 473 \\%& 5695 \\
$D^{\ast+} \rightarrow (K^-\mu^+\nu)\pi^+$          
&  3  &  3  &  4  & 412 \\%& 4576 \\
$D^{\ast+} \rightarrow (K^-e^+\nu)\pi^+$            
&  3  &  3  &  3  & 276 \\%& 3684 \\
$D^{\ast+} \rightarrow (K^-\pi^+(\pi^0))\pi^+$      
&  5  &  5  &  5  &1475 \\%&15135 \\
$D^{+}~ \rightarrow K^-\pi^+\pi^+$                  
&  5  &  4  &  4  &1058 \\%&12086\\
$D^0~~ \rightarrow K^-\pi^+$                        
&  4  &  5  &  4  &1076 \\%&12811\\
$D^0~~ \rightarrow K^-\pi^+(\pi^0)$                 
&  5  &  5  &  6  &1947 \\%&21407\\
      \hline
    \end{tabular}
  \end{center}
  \caption[]{Number of bins in each dimension used for the individual
           decay modes and the  number of data events per bin
           at $\sqrt{s} = 92.990$\,GeV.}
  \label{mbinning}
\end{table}

%=========================================================================%
\subsection{\label{likelihood} The maximum likelihood fit}
%=========================================================================%
The data taken by the DELPHI detector in the years 1993 and 1995 at energies
near the $Z$ resonance allow the investigation of the energy dependence of the
forward--backward asymmetry of $c$ and $b$ quarks. Due to the reduced
statistics of 182386 events at
$\sqrt{s} = 89.434$\,GeV and 265877 events at $\sqrt{s} = 92.990$\,GeV, a
binned maximum likelihood fit is used instead of the $\chi^2$ fit
of section \ref{chi2}.
The likelihood function is given by:
\begin{equation}
{\cal L} = \sum\limits_{i=1}^{N_{bins}}
 \left\{ 
  \ln \frac { \lambda_{i}^{+ ^{N_{i}^{+}} } } { e^{\lambda_{i}^{+}}\cdot N_{i}^{+}! } +
  \ln \frac { \lambda_{i}^{-^{N_{i}^{-}}} } { e^{\lambda_{i}^{-}}\cdot N_{i}^{-}! }
 \right\}
\end{equation}
where the $N_{bins}$ cells are given by the binned information in $X_E$, 
$tr({\cal P}_{ev})$ and $\cos\theta_{thrust}$. The bins for the three dimensions are given in tables \ref{pbinning} and \ref{mbinning}.
The $\lambda_{i}^{\pm}$ describe the expectation in each cell of
a Poisson distribution, and are given by:
\begin{equation}\label{like-lam-eqn}
\lambda_{i}^{\pm} = \frac {N_i^{tot}} {2} \sum\limits_{j=1}^{7}
                    f_{ij} C_{ij} (1 \pm A_{FB}^j (\cos\theta_i)).
\end{equation}
The total number of candidates $N_i^{tot} = N_i^+ + N_i^-$ is taken from real data and
the coefficients $f_{ij}$, $C_{ij}$ and $A_{FB}^j (\cos\theta_i)$ are defined 
in the same way as in section \ref{chi2}.

%=========================================================================%
\section{\label{mixing} Effective mixing in {\boldmath $b \to D$} decays}
%=========================================================================%

The observed forward--backward asymmetry in $b$ events is proportional to the
charge correlation of the reconstructed $D$ meson to the primary quark. The
correlation $C_b$ is reduced because of two factors. The first is  
$B^0-\bar B^0$ mixing, and the second is double $D$ production in $B$ decays. 
The observed $b$ asymmetry in the different $D$ samples
needs to be corrected for both of the above sources of $D$ mesons of the wrong
charge state. 
%The effective magnitude of these effects
%depends on details of the $B \to D$ and $B \to {\bar D}$ decay properties.  

The  effect of $B^0-\bar B^0$ mixing is to reduce the correlation $C_b$
by a factor
$1-2\chi$. The mixing probability $\chi$ is determined by the
mass difference $\Delta m$ between the two mass eigenstates and by the
$B$ lifetime $\tau_B$. The product of these leads to  
$\chi_d = 0.172 \pm 0.010$ \cite{PDG98}. For the $B^0_s$ only a lower limit of
$\Delta m_s > 14 \, {\rm ps}^{-1} $ \cite{PDG98}
is known. This is compatible with full mixing $\chi_s \sim 0.5$.

The production of $D$ mesons from the ``upper vertex" (i.e. via the $W$ decay,
rather than from the $bcW$ coupling) 
also reduces
the charge correlation to the primary $b$ quark. A sizeable rate
$f_{B\to W\to D}$ of wrong sign $D$ mesons from $B$ decays reduces
the measured $b$ asymmetry by a factor $1-2\cdot f_{B\to W\to D}$.   
Recent measurements of CLEO and ALEPH indicate a significant rate of
double-charmed $B$ decays involving no $D_s$ production
($D_s$ production would not result in $D$ or  $D^{*}$ of the wrong charge
state). 
CLEO \cite{CLEODD} finds for a mixture of $B^0_d$ and $B^+$ a ratio of:

\begin{equation}
  \frac{\Gamma(B \to DX)}{\Gamma(B \to \bar{D}X)} = 
0.100\,\,\pm\,\,0.026\,(stat)\,\,\pm\,\,0.016 \,(syst) .
\end{equation}

\noindent
This number is in good agreement with the ALEPH result on double $D$ production
in $B$ decays \cite{ALEPHDD}:

\begin{equation}
  BR(b \to D^0\bar{D}^0,D^0D^-,D^+\bar{D}^0 \,+\, X) =
 \left (7.8 \, {^{+2.0}_{-1.8}} \, {^{+1.7}_{-1.5}} \, {^{+0.5}_{-0.4}}\right ) \, \%,
\end{equation}

\noindent
where the dominant contribution is $B \to D^{(*)}\bar{D}^{(*)}K$. The first
error is statistical, the second one contains the systematic errors and the
third  corresponds to the uncertainties on the $D$ branching fractions.
Both
measurements include Cabbibo suppressed $W \to c\bar d$ decays which are
expected to contribute with a rate of about 1\% per $W$ decay to the 
``upper vertex" charm rate.

Mixing is relevant for $D$ mesons from $B^0_d$ and $B^0_s$ decays, but not
from $B^+$ or $\Lambda_b$ decays, whereas the effect from the ``upper vertex"
charm contributes in all $B$ decays. The fractions of $D^\pm$, \dnb\ and
$D^{\ast \pm}$ from different $B$ states need to be determined from the
branching rates $B \to D$ and $B \to {\bar D}$ to be able to consider the
combination of both
effects
correctly. Very little is known at 
present about the
individual exclusive branching ratios, but several
inclusive measurements from the $\Upsilon(4S)$ and LEP experiments can be used
to deduce the rates.

CLEO and ARGUS have measured the rates \cite{PDG98} of \dnb , $D^\pm$, $D^\pm_s$
and $\Lambda_c^\pm$ as well as the rates of $D^{\ast \pm}$ and \dnsb\
in $\Upsilon(4S)$ decays, i.e. in decays of $B^0_d$ and $B^+$ at about 50\,\%
admixture. From these measurements the overall rates of
$B^0_d$ and $B^+$ decays in the
\db\ samples are deduced, taking into account the production fractions of
$B^+$, $B^0_d$, $B^0_s$ and $\Lambda_b$ in $Z \to b\bar b$ events \cite{PDG98}.
The relative rate of $D^{*\pm}$ from $B^-$ and from $\bar{B}^0$ is not measured. Therefore the
JETSET prediction of 
\begin{equation}
(B^+\to D^{*\pm}\,X) / (B^{+,0} \to D^{*\pm}\,X) = 0.30
\end{equation}
with a relative error of 50\,\% is used. This number is compatible with the
assumption that most of the $B^+\to D^{*\pm}\,X$ decays are produced via 
$D^{\ast\ast}$ (i.e. $D_2^\ast,D_1^\ast,D_1,D_0$) decays and higher $D$
resonances.

The $D^\ast$ rates measured at the $\Upsilon(4S)$ also fix the effective rate
of vector and pseudoscalar mesons $V/(V+P)$.
The decay of the $D^{\ast0}$ into $D^+\pi^-$ is forbidden by phase space, while
the branching ratio $D^{\ast+} \to D^0\pi^+$ is measured to be
$0.683 \pm 0.013$ \cite{PDG98}.
This difference in 
%LLLLLLLLL My next line instead of yours 2 lines down
charged and neutral $D^{\ast}$
%$D^{\ast+}$ and $D^{\ast0}$ 
decays  significantly affects the
fractions of $B^+$ and $B^0_d$ decays seen in
the $D^{\pm}$ and \dnb\ samples and thus the effective mixing.

A small correction to the $B^+$ and $B^0_d$ into $D^\pm$, \dnb\ and $D^{\ast \pm}$
rates originates from $D^{\ast\ast}$ production. These states 
subsequently decay into vector or pseudoscalar charm mesons. $D^\ast_2$
mesons decay into $D^\ast$ and $D$ states and the ratio between these two
decays is given by the measurement $BR(D^{\ast0}_2 \to D^+\pi^-)/BR(D^{\ast0}_2 \to D^{\ast+}\pi^-)= 2.3 \pm 0.8$ \cite{D** CLEO ARGUS}.
Angular momentum conservation
allows $D^{\ast}_1$ and $D_1$ mesons to decay into
$D^{\ast}$ states, and $D_0$ to decay into $D$ states.
The decay rates of the different $D^{\ast\ast}$ states into charged and
neutral states ($D^{\ast}\pi$,$D\pi$) are fixed by isospin invariance.
The relative production rates of the
four $D^{\ast\ast}$ states are assumed to be proportional to the number of
spin states. The total $D^{\ast\ast}$ rate is obtained from the
measured semileptonic $B$ branching ratios 
$BR(B \to \bar D^{\ast\ast}l^+\nu)/BR(B \to Xl^+\nu) = 0.26 \pm 0.07$ \cite{PDG98}.

The decay of $B^0_s$ and $\Lambda_b$ into $D^\pm$, \dnb\ and $D^{\ast \pm}$
also contribute to the sample. They can be deduced from the total rate of
$D^0$, $D^+$, $D^+_s$ and $\Lambda_c^+$ in $Z \to b\bar b$ events
measured at LEP \cite{LEP b->D}. Here the number of charm quarks produced per
$b$ decay is limited to the LEP measurement $n_c = 1.17 \pm 0.04$
\cite{PDG98}, with charmonia and $\Xi_c$ production taken into account.

From this information and assuming that the branching fraction of $W\to c\bar s\to D^{(\ast)+}$ is
equal to that for $W\to c\bar s\to D^{(\ast)0}$, the effective $B$ mixings
in the $D^+$, $D^0$ and $D^{\ast+}$ samples are:
\begin{eqnarray}\nonumber
\chi_{eff}(D^+)       &=& 0.222 \pm 0.033 \\
\chi_{eff}(D^0)       &=& 0.176 \pm 0.030 \\\nonumber
\chi_{eff}(D^{\ast+}) &=& 0.222 \pm 0.033\,\, .
\end{eqnarray}

\noindent
The errors quoted represent the precision of
the measurements used to determine the $b$ decay properties. The effective
mixing for the $D^{\ast+}$ sample is in good agreement with a direct
measurement of OPAL using the jet charge technique in the
hemisphere opposite to the reconstructed meson. They obtained
$\chi_{eff}(D^{\ast+}) = 0.191 \pm 0.083$ \cite{OPAL Afb D}.

%=========================================================================%
\section{\label{sec-qcd} QCD corrections}
%=========================================================================%

The analysis of the final state of hadronic $Z$ decays gives only indirect
information about the electroweak process $Z \to q\bar{q}$. The evolution
to the final parton level and the following process of hadronization smear the
clear signature of the initial $q\bar{q}$ system. The hard gluon radiation
in the parton shower changes significantly the direction of the primary quarks
and thus also the angular distribution of the following partons and hadrons.
The size of this QCD effect strongly depends on the individual techniques in
the determination of 
the forward--backward asymmetry.
The QCD correction can be written as \cite{HF-QCD}:
\begin{equation}
\Afbq = (1-C_{q})\, (\Afbq)_0
       = (1-s_q\,C_{\rm QCD}^q)\, (\Afbq)_0
\end{equation}
where $(\Afbq)_0$ is the asymmetry without gluon radiation,
which can be calculated from the measured asymmetry \Afbq\ through the
correction coefficient $C_{q}$. This correction coefficient can be
parameterized by a bias factor $s_q$, which accounts for the individual
sensitivity to the QCD correction $C_{\rm QCD}^q$. The values of the QCD
corrections are estimated to be \cite{HF-QCD}:
\begin{eqnarray}\nonumber
C_{\rm QCD}^b &=& (2.96\,\,\pm\,\,0.40)\,\% \\
C_{\rm QCD}^c &=& (3.57\,\,\pm\,\,0.76)\,\%\,\,.
\end{eqnarray}

\noindent
%The asymmetry fit is performed using the initial quark axis in the simulation (see
%equation \ref{chi2-eqn} and \ref{like-lam-eqn}),
%therefore
%most of the QCD corrections  are implicitly taken into account.
The experimental bias is studied using a fit to the simulation after setting
the generated asymmetry to 75\,\%. The observed relative difference $C_q$ of 
$(-1.66 \pm 0.01)\,\%$ for $c\bar c$ and $(-2.22 \pm 0.02)\,\%$ for $b\bar b$
lead to the bias factors $s_c = -46\,\%$ and $s_b = -75\,\%$.
The values of $C_q$ are 
used to define the experimental
bias on the QCD corrections to correct the fit results.
The statistical error of the fit and the
uncertainty of the QCD correction are used to determine the systematic error.

%=========================================================================%
\section{Systematic uncertainties}
%=========================================================================%

\begin{table}[p]
\renewcommand{\arraystretch}{1.15}
\begin{center}
\begin{tabular}{|c||r@{.}l@{$~\pm~$}l||c|c||c|c||c|c|} 
\hline

  systematic                      &  \multicolumn{3}{c||}{} &
\multicolumn{2}{c||}{$91.235$\,GeV}&
\multicolumn{2}{c||}{$89.434$\,GeV}&
\multicolumn{2}{c|}{$92.990$\,GeV}\\

error source & \multicolumn{3}{c||}{variation} & 
 $\delta \Afbc$  & $\delta \Afbb$ &
 $\delta \Afbc$  & $\delta \Afbb$ &
 $\delta \Afbc$  & $\delta \Afbb$ \\

  & \multicolumn{3}{c||}{} & 
 $\times 10^3$ &  $\times 10^3$ & 
 $\times 10^3$ &  $\times 10^3$ & 
 $\times 10^3$ &  $\times 10^3$ \\
\hline

MC statistics
 & \multicolumn{3}{c||}{see text} 
 &  $\pm 2.47$  &  $\mp 3.54$  
 &  $\pm 3.31$  &  $\mp 7.20$  
 &  $\pm 3.31$  &  $\mp 7.20$  \\ \hline

$\langle X_E \rangle_{D^\ast}$                       
 & 0&510 & 0.009  
 & $\pm 0.16$  &  $\mp 0.40$  
 & $\mp 0.11$  &  $\mp 0.36$       
 & $\pm 0.59$  &  $\mp 0.60$     \\    

$\langle X_E \rangle_B$                       
 & 0&702 & 0.008  
 &  $\mp 0.20$  &  $\pm 0.25$  
 &  $\pm 0.69$  &  $\mp 0.51$       
 &  $\mp 0.25$  &  $\mp 0.43$     \\    

$\epsilon_{B \to D}$                       
 & 0&42 & 0.07  
 &  $\pm 0.26$  &  $\mp 0.40$
 &  $\mp 0.78$  &  $\mp 0.18$        
 &  $\pm 0.42$  &  $\pm 0.98$     \\    

$\tau(B^+)$                              
% & \multicolumn{3}{c||}{see text} 
 & 1&65 & 0.04  
 &  $\mp 0.06$  &  $\pm 0.18$  
 &  $\mp 0.02$  &  $\pm 0.31$           
 &  $\mp 0.10$  &  $\pm 0.49$     \\    

$\tau(B^0)$                              
% & \multicolumn{3}{c||}{see text} 
 & 1&56 & 0.04  
 &  $\mp 0.01$  &  $\mp 0.69$  
 &  $\pm 0.09$  &  $\mp 0.20$        
 &  $\mp 0.11$  &  $\mp 0.28$     \\    

$\tau(B^0_s)$                              
% & \multicolumn{3}{c||}{see text} 
 & 1&54 & 0.07  
 &  $\mp 0.01$  &  $\mp 0.03$  
 &  $\pm 0.02$  &  $\pm 0.04$        
 &  $\mp 0.04$  &  $\pm 0.08$     \\    

$\tau(\Lambda_b)$                              
% & \multicolumn{3}{c||}{see text} 
 & 1&22 & 0.05  
 &  $\mp 0.03$  &  $\pm 0.11$  
 &  $\pm 0.08$  &  $\mp 0.02$        
 &  $\mp 0.08$  &  $\pm 0.11$     \\    

$\tau(D^+)$                              
% & \multicolumn{3}{c||}{see text} 
 & 1&057& 0.015 
 &  $\pm 0.02$  &  $\mp 0.03$  
 &  $\pm 0.01$  &  $\mp 0.09$         
 &  $\pm 0.02$  &  $\mp 0.03$     \\    

$\tau(D^0)$                              
% & \multicolumn{3}{c||}{see text} 
 & 0&415& 0.004 
 &  $\pm 0.04$  &  $\pm 0.02$  
 &  $\mp 0.25$  &  $\pm 0.10$         
 &  $\pm 0.03$  &  $\pm 0.13$     \\    

$\tau(D^+_s)$                              
% & \multicolumn{3}{c||}{see text} 
 & 0&467& 0.017 
 &  $\mp 0.02$  &  $\pm 0.04$  
 &  $\mp 0.10$  &  $\pm 0.11$         
 &  $\mp 0.08$  &  $\pm 0.11$     \\    

$\tau(\Lambda_c)$                              
% & \multicolumn{3}{c||}{see text} 
 & 0&206& 0.012 
 &  $\pm 0.03$  &  $\mp 0.04$  
 &  $\mp 0.03$  &  $\pm 0.04$         
 &  $\mp 0.03$  &  $\pm 0.04$     \\    

$f(D^+)$      
 & 0&221 & 0.020
 & $\mp 0.02$  & $\pm 0.04$ 
 & $\mp 0.35$  & $\pm 0.46$ 
 & $\mp 0.27$  & $\pm 0.36$ \\

$f(D^+_s)$    
 & 0&112 & 0.027
 & $\pm 0.19$  & $\mp 0.19$ 
 & $\mp 0.42$  & $\pm 0.42$ 
 & $\pm 0.13$  & $\mp 0.19$ \\

$f(c_{baryon})$
 & 0&084 & 0.022
 & $\pm 0.03$  & $\mp 0.03$ 
 & $\mp 0.23$  & $\mp 0.05$ 
 & $\pm 0.20$  & $\pm 0.15$ \\ 

$n_{g\to c\bar c}$ 
 & 2&38 &0.48 \% 
 & $\pm 0.05$  & $\pm 0.19$  
 & $\mp 0.05$  & $\pm 0.19$  
 & $\pm 0.05$  & $\pm 0.19$  \\

%$(R_b\cdot P_{b\to D})/(R_c\cdot P_{c\to D})$   
$\frac{(R_b\cdot P_{b\to D})}{(R_c\cdot P_{c\to D})}$   
 & \multicolumn{3}{c||}{see text} 
 &  $\pm 0.38$  &  $\mp 0.40$  
 &  $\mp 1.15$  &  $\pm 0.07$  
 &  $\pm 0.70$  &  $\pm 0.22$  \\

eff. mixing            
 & \multicolumn{3}{c||}{see text} 
 &  $\mp 0.05$  &  $\pm 5.78$  
 &  $\mp 0.26$  &  $\pm 2.82$  
 &  $\mp 0.18$  &  $\pm 5.71$  \\

QCD bias
 & \multicolumn{3}{c||}{see text} 
 &  $\mp 0.24$  &  $\mp 0.23$  
 &  $\mp 0.17$  &  $\mp 0.18$  
 &  $\mp 0.41$  &  $\mp 0.27$  \\ \hline

fit method                                       
 & \multicolumn{3}{c||}{see text} 
 &  $\pm 1.73$  &  $\mp 2.81$  
 &  $\pm 1.73$  &  $\mp 2.81$  
 &  $\pm 1.73$  &  $\mp 2.81$  \\

$b_{tag}$                                       
 & \multicolumn{3}{c||}{see text} 
 &  $\pm 0.74$  &  $\pm 1.54$  
 &  $\pm 0.74$  &  $\pm 1.54$  
 &  $\pm 0.74$  &  $\pm 1.54$  \\

$R_{S/B}$
 & \multicolumn{2}{r@{~$\pm$~}}{} & \multicolumn{1}{r||}{$5/10 \%$}  
 &  $\mp 0.63$  &  $\mp 1.05$  
 &  $\pm 1.81$  &  $\mp 1.28$  
 &  $\mp 2.51$  &  $\mp 2.54$  \\

%$\pi_{slow}$,wrong sign                          
% &  \multicolumn{2}{r@{~$\pm$~}}{} & \multicolumn{1}{r||}{$30 \%$} 
% &  $\pm 0.41$  &  $\pm 0.89$  
% &  $\mp 2.15$  &  $\pm 2.02$  
% &  $\pm 0.96$  &  $\pm 1.75$  \\

$\pi_{slow}$,wrong sign                          
 &  \multicolumn{2}{r@{~$\pm$~}}{} & \multicolumn{1}{r||}{$15 \%$} 
 &  $\pm 0.21$  &  $\pm 0.43$  
 &  $\mp 1.07$  &  $\pm 0.97$  
 &  $\pm 0.48$  &  $\pm 0.85$  \\

$A^{uds}_{FB}(backgr.)$                             
 &  \multicolumn{2}{r@{~$\pm$~}}{} & \multicolumn{1}{r||}{$30 \%$}    
 &  $\mp 0.55$  &  $\pm 0.02$  
 &  $\mp 1.46$  &  $\mp 0.50$  
 &  $\mp 0.69$  &  $\mp 0.54$  \\

$A^{b,c}_{FB}(backgr.)$                             
 &  \multicolumn{2}{r@{~$\pm$~}}{} & \multicolumn{1}{r||}{$30 \%$}  
 &  $\mp 1.26$  &  $\mp 3.62$  
 &  $\pm 2.04$  &  $\mp 7.86$  
 &  $\mp 3.93$  &  $\mp 6.59$  \\
\hline\hline
total &\multicolumn{3}{c||}{---}   
 & $\pm 3.51$ & $\mp 8.47$ 
 & $\pm 5.28$ & $\mp11.67$ 
 & $\pm 6.16$ & $\mp12.20$ \\
%\hline\hline
%total &\multicolumn{3}{c||}{---}   
% & $\pm 2.50$ & $\mp 7.70$ 
% & $\pm 4.12$ & $\mp 9.19$ 
% & $\pm 5.20$ & $\mp 9.85$ \\
\hline
\end{tabular}
\end{center}
\caption{\label{systematics} Contributions to the systematic errors on the
measured asymmetries.}
\end{table}

The systematic error sources are of two types. The uncertainty of the
simulation modelling of heavy quark production affects the measurement, and 
the fit method
itself is a potential source for a systematic error.

To describe adequately  heavy quark production, it is necessary to correct for 
inadequate simulation settings. This is achieved using
JETSET; the relevant distributions are compared (at the full simulation level,
but before detector acceptance effects) for the parameters as used in the
generation and at their required values.
%to produce the required distribution and compare it to the one given
%in the full detector Monte Carlo before detector acceptance. 
The ratio of the two
spectra is used as a weight to modify the simulation shape in
equations \ref{chi2-eqn} and \ref{like-lam-eqn}. To estimate the systematic
uncertainty, the input value
is changed within its error and the procedure is repeated.

A similar approach is employed to allow for the uncertainty of the means 
$\langle X_E^c(D)\rangle$ and $\langle X_E^b(B)\rangle$.
%, the procedure is quite similar.
JETSET is used to generate
the $X_E$ distributions of all charm states according to 
$\langle X_E^c(D^{\ast}) \rangle =0.510 \pm 0.005 \pm 0.008$, 
$\langle X_E^b(B) \rangle =0.702 \pm 0.008$~\cite{LEPHF}.
The energy spectrum of $D$ mesons in the $B$ rest frame 
was measured by CLEO~\cite{cleo}. This spectrum
includes the contributions from
$B \to \bar{D} \, X$ and $B \to D\bar D \, X$. It can be parameterized in terms of a
Peterson function with $\epsilon_{b \to D} = 0.42 \pm 0.07$~\cite{LEPHF}.

The corrections are applied to all simulated charm ground state
hadrons separately
for $b\bar b$ and $c\bar c$ events. The resulting $X_E$ distribution of
the sum of all charm hadron ground states in $c\bar c$ events is found to be 
in agreement with the corresponding
average of $\langle X_E^c(D^0,D^+) \rangle=0.484 \pm 0.008$~\cite{LEPHF}.
Here the effect of gluon splitting into $c\bar{c}$ is taken into account.
The systematic uncertainties
are calculated separately for $\langle X_E^c(D)\rangle$,
$\langle X_E^b(B)\rangle$ and $\epsilon_{b \to D}$.

The $b$--hadron lifetimes are corrected separately for $B^+$, $B^0$,
$\Lambda_b$ and $B^0_s$. Here the world averages 
$\tau(B^0)=1.56\pm0.04$, $\tau(B^+)=1.65\pm0.04$,
$\tau(B^0_s)=1.54\pm0.07$ and $\tau(\Lambda_b)=1.22\pm0.05$ ps
\cite{PDG98} are used to correct the
simulation. For the systematic uncertainties from this
source, all the $b$ lifetime distributions are regenerated with
a change of one standard deviation  and the fit is performed again.
Similarly the $c$--hadron lifetimes are also corrected
separately for $D^+$, $D^0$, $\Lambda_c$ and $D^+_s$.
Here the values $\tau(D^0)=0.415\pm0.004$,
$\tau(D^+)=1.057\pm0.015$, $\tau(D^+_s)=0.467\pm0.017$ and
$\tau(\Lambda_c)=0.206\pm0.012$ ps from~\cite{PDG98} are taken.

The separation between $b\bar b$ and $c\bar c$ events obtained from 
the $b$--tagging depends on the production rates of $D^+$ and $D^0$ 
mesons
in $c\bar c$ events. The rates of charm hadrons in the hemisphere
opposite to the reconstructed $D$ are therefore fixed to the present
averages $f(D^+)=0.221\pm0.020$, $f(D^+_s)=0.112\pm0.027$ and
$f(c_{baryon})=0.084\pm0.022$~\cite{EWWPPE}. The $D^0$ rate is calculated from
these numbers according to:
\begin{eqnarray}
f(D^0) = 1 - f(D^+) - f(D^+_s) - f(c_{baryon}) \, .
\end{eqnarray}

\noindent 
A variation of one standard deviation on each fraction is included in the
systematic error, leaving the $D^0$ fraction free to keep the sum constant.

The effect due to the efficiency of the $b$-tagging was studied in
reference \cite{delphi_rb} using a tuning determined independently on data and 
simulation. A residual difference in the $b$
efficiency of 3\,\% per jet between data and simulation was found.
The corrections to the physics parameters in the simulation mentioned
above account for this difference. This systematic error is therefore
already included in the physics corrections.
Furthermore, the effect due to the resolution of the $b$-tagging is
determined by interchanging the $b$-tag tunings of data and simulation
with each other.

The rate of gluon splittings into $c\bar c$ pairs is varied by one
standard deviation.

The relative rate of $D$ mesons from $b\bar b$ and $c\bar c$ events is not a
free parameter in the asymmetry fit. Therefore the ratio is fixed to the
DELPHI measurement \cite{DELPHI b->D} using the same data and varied by
one standard deviation.

The mixing correction for  \Afbb ~is discussed in section \ref{mixing}.
The systematic error is obtained by varying separately each parameter
used to obtain the $B$ decay rates into \db\ mesons by one standard deviation. 
The effect of the variation is studied
directly on the asymmetry fit; this allows for the
lifetime dependence of the $B^0 - \bar B^0$ mixing.
The total error is then calculated taking the correlation between the
parameters into account. The effect of the oscillation frequency error is small
compared to that from the uncertainty of the $B \to \db$ rates.

Differences between the signal and background efficiency as a function of 
$\cos\theta$ are considered in the calculation of the probabilities from the
simulation. 
%LLLmmLLLLLLLLLLL     I've changed next sentence
Because the asymmetry evaluation depends on the ratio of $D$ to $\bar D$
at a given $\cos\theta$,
%Since the asymmetry enters in the $\chi^2$ as a
%function of $\cos\theta$, 
the sensitivity to efficiency variations (which are largely independent of the
nature of the $D$) is small.
The systematic error due to the fit method is estimated by 
comparing the results of the $\chi^2$ fit to the results 
obtained assuming a Poisson distribution and neglecting in both cases
the error on the simulation. The observed difference is included
in the systematic error.

For all decay modes the relative normalization $R_{S/B}$ is obtained from
a fit of the simulated $D$ signal and background to the data. A 
variation of $\pm 5\,\%$ ($\pm 10\,\%$ for the off--peak data) is included in the systematic error, not only to
account for the error of the fitted $R_{S/B}$, but also for uncertainties
in the agreement of the shape of the mass difference signals in data and
simulation.

The contribution of partially reconstructed $D$ decays depends on the
efficiency to reconstruct such $\pi_{sl} + X$ combinations, as well as
on the total rate of $D^{\ast+} \rightarrow D^0\pi^+$ decays in hadronic
$Z$ events.
The differences for the background normalizations between data and simulation 
average around $10\,\%$ whereas the total rate of $D^{\ast+}
 \rightarrow D^0\pi^+$ decays is known at the $5\,\%$ level. 
Combining these, the contribution to the systematic error
is estimated as a $\pm 15\,\%$ variation
of the prediction of the simulation. 

The three classes of the combinatorial background ($uds$, $c$ and $b$)
have a small remaining asymmetry.
The asymmetry of the $uds$ quark background is taken from the simulation
and subtracted in the fit.
The charge correlation for the combinatorial
background from $b$ and $c$ events is taken from the simulation.
The agreement between data and simulation is tested
using the side bands, where good agreement is found. 
$30\,\%$ of the effect due to the background asymmetry is considered in the
systematic error on the asymmetry.

%For almost all sources the linearity of the systematic uncertainty is checked
%by varying the parameters over a much wider range than the given one. For all variations a good
%linearity was found which displays the absence of any further problems.

The contributions to the systematic errors for the combined fit of the charm
and bottom asymmetries are listed in table
\ref{systematics}. The relative sign of the systematic error indicates the
direction in which the results change for a particular error source.

%=========================================================================%
\section{Fit results}
%=========================================================================%

%--------------------------------- asymmetry table 1 --------------------------

The results of the 2 parameter fits of the $c$ and $b$ asymmetries for
the different energies are given
in tables \ref{asymmetries},  \ref{m2asymmetries} and \ref{p2asymmetries}. 

\begin{figure}[p]
 \begin{center}
   \mbox{\epsfig{file=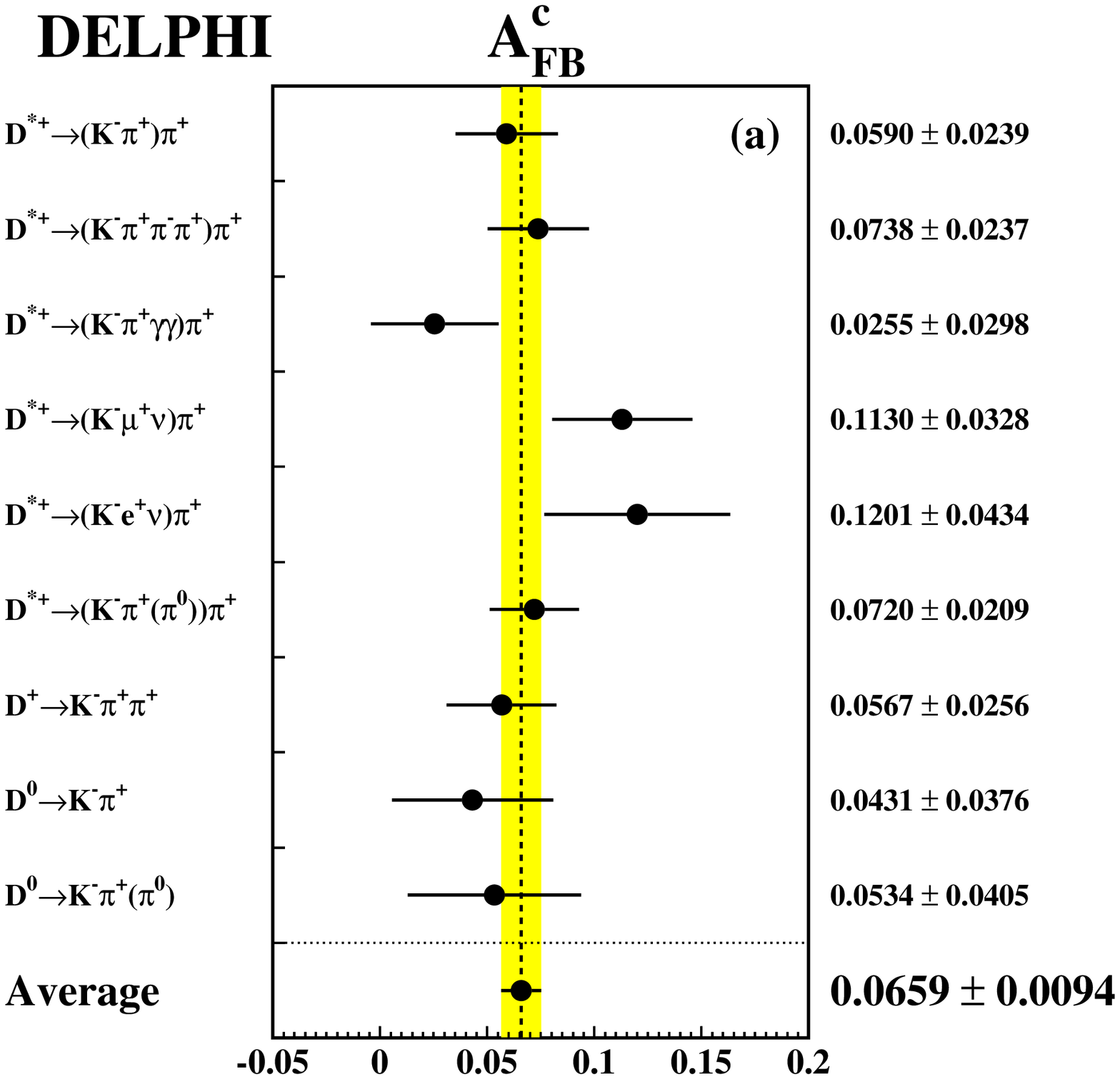,width=10cm}} \\
   \mbox{\epsfig{file=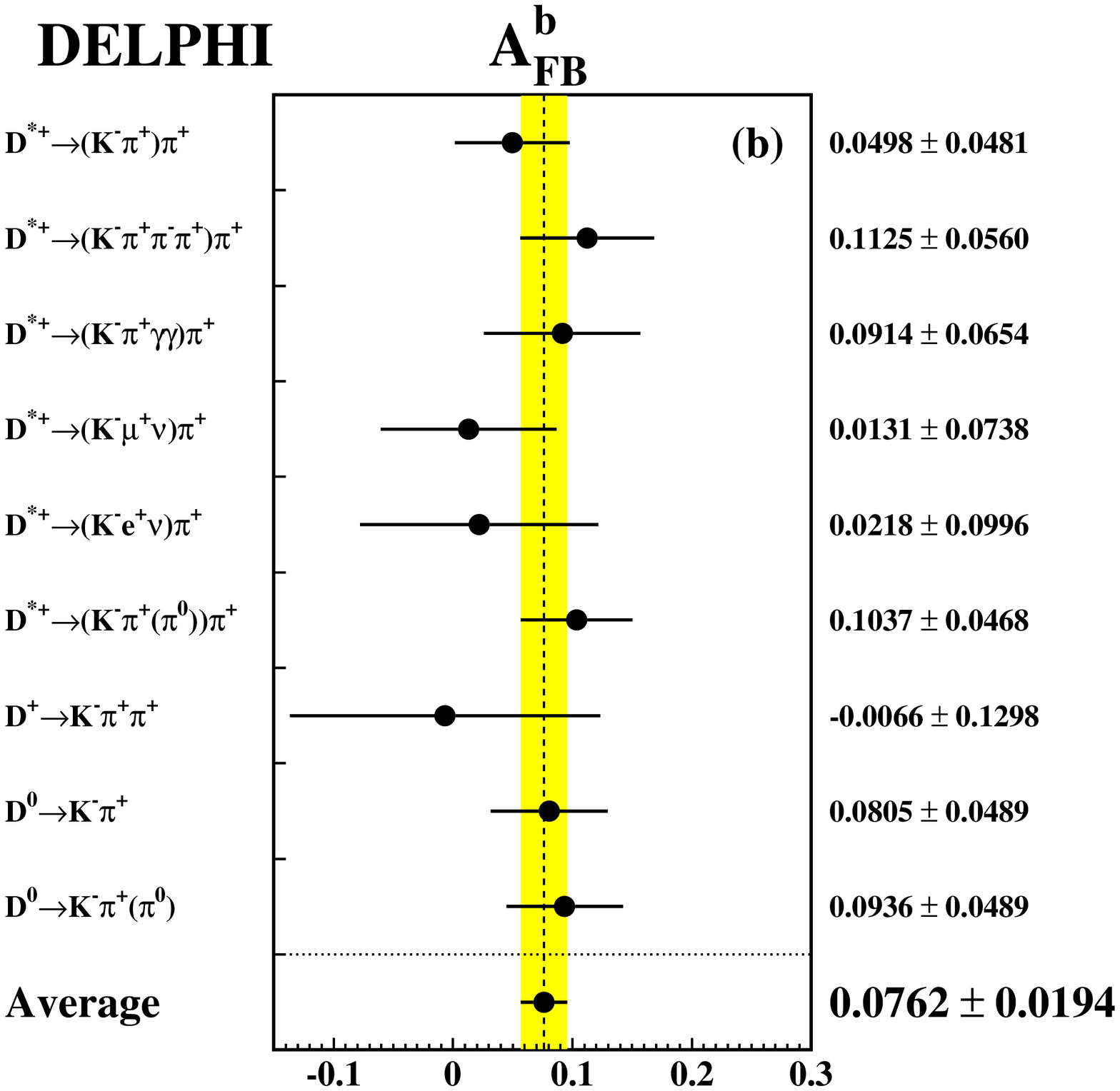,width=10cm}}
 \end{center}
 \caption[]{\label{fit_res} The results of the two parameter fit of 
           the $c$ and $b$ asymmetry at an average centre--of--mass
           energy of 91.235\,GeV
           for the different $D$ samples are shown.
           The grey bands represent the averages
           over all these measurements. Only statistical errors are shown.}
\end{figure}

\begin{figure}[p]
 \begin{center}
  \mbox{\epsfig{file=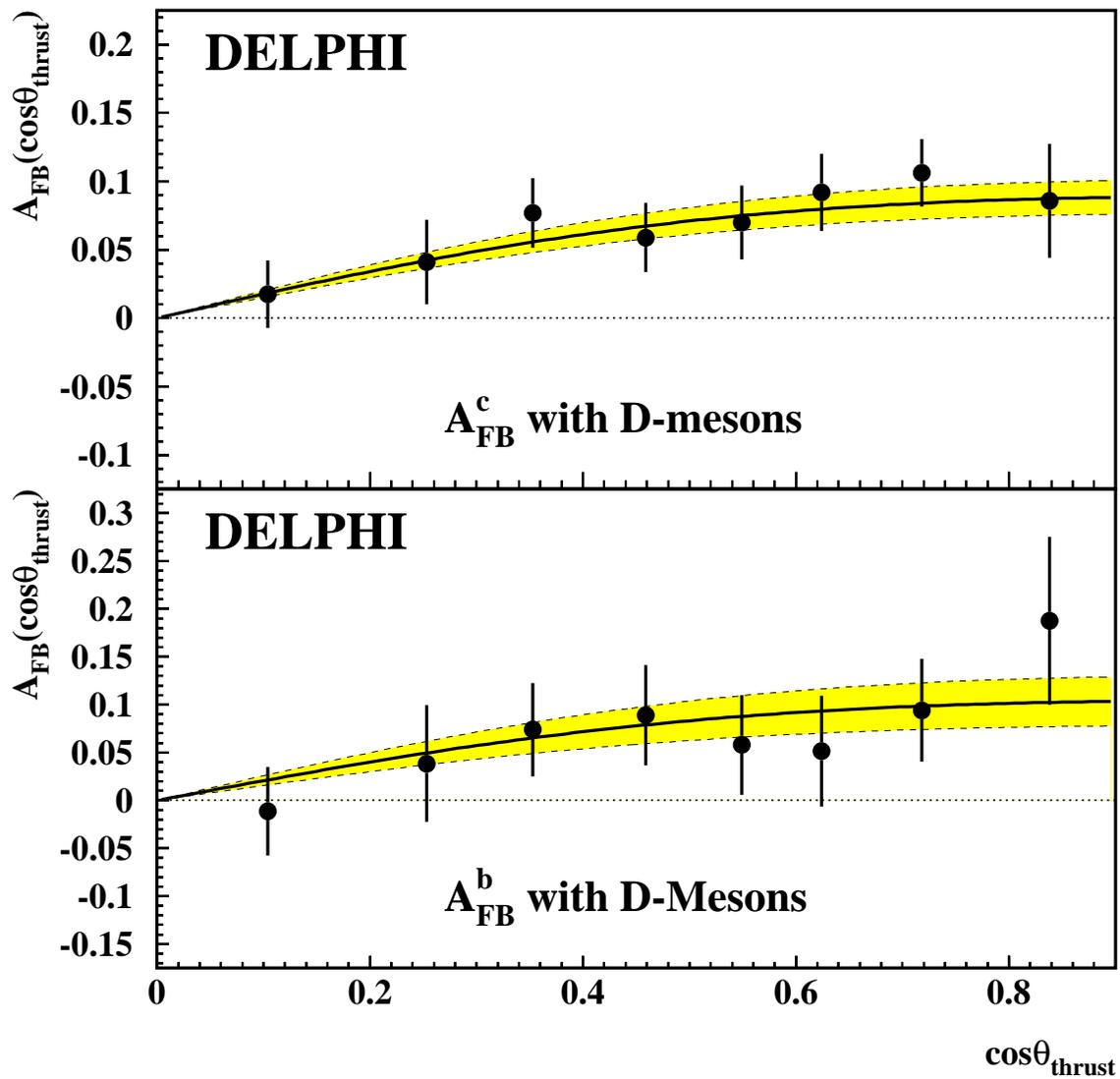,width=17cm}}
 \end{center}
\caption[]{\label{diffasy} The $c$ and $b$ forward--backward asymmetries
           at an average centre--of--mass energy of 91.235\,GeV
           as a function of $\cos\theta_{thrust}$. Only statistical errors are shown.
           The bands represent the fit results.}
\end{figure}

Taking the correlations into account, the combination  of the results of the
different samples at the peak energy leads to:
\[
    \renewcommand{\arraystretch}{1.6}
\begin{array}{rcl}
   \Afbc(\sqrt{s} = 91.235\,{\rm GeV}) 
        &=& 0.0659\,\, \pm\,\, 0.0094\, (stat) \\
   \Afbb (\sqrt{s} = 91.235\,{\rm GeV}) 
        &=& 0.0762\,\, \pm\,\, 0.0194\, (stat) \\
\end{array}
\]

\noindent
with a statistical correlation of $-0.27$. The average centre--of--mass
energy is $\sqrt{s} = 91.235$ GeV. 
In figure \ref{fit_res}
the fit results for the forward--backward asymmetries of the different samples 
are compared to the average. The forward--backward asymmetry averaged over all
samples as a function of $\cos\theta_{thrust}$ is shown in figure \ref{diffasy}.
\begin{table}[h]
    \renewcommand{\arraystretch}{1.15}
  \begin{center}
    \begin{tabular}{|l|rcr|rcr|c|c|}
       \hline
~~~~~~~decay mode         &               
\multicolumn{3}{c|}{$A_{FB}^{c}$} &
\multicolumn{3}{c|}{$A_{FB}^{b}$} & correlation & $\chi^2/N.D.F.$   \\
\hline
$D^{\ast+} \rightarrow (K^-\pi^+)\pi^+$   & 
$0.0590$ & $\pm$ & $0.0239$ & $0.0498$ & $\pm$ & $0.0481$ & $-0.26$ & $0.97$ \\

$D^{\ast+} \rightarrow (K^-\pi^+\pi^-\pi^+)\pi^+$ & 
$0.0738$ & $\pm$ & $0.0237$ & $0.1125$ & $\pm$ & $0.0560$ & $-0.28$ & $0.87$ \\

$D^{\ast+} \rightarrow (K^-\pi^+\gamma\gamma)\pi^+$ & 
$0.0255$ & $\pm$ & $0.0298$ & $0.0914$ & $\pm$ & $0.0654$ & $-0.28$ & $0.79$ \\

$D^{\ast+} \rightarrow (K^-\mu^+\nu)\pi^+$ &
$0.1130$ & $\pm$ & $0.0328$ & $0.0131$ & $\pm$ & $0.0738$ & $-0.26$ & $1.10$ \\

$D^{\ast+} \rightarrow (K^-e^+\nu)\pi^+$ &
$0.1201$ & $\pm$ & $0.0434$ & $0.0218$ & $\pm$ & $0.0996$ & $-0.23$ & $1.26$ \\

$D^{\ast+}\rightarrow (K^-\pi^+(\pi^0))\pi^+$ & 
$0.0720$ & $\pm$ & $0.0209$ & $0.1037$ & $\pm$ & $0.0468$ & $-0.27$ & $0.92$ \\

$D^+~ \rightarrow K^-\pi^+\pi^+$  & 
$0.0567$ & $\pm$ & $0.0256$ & $-0.0066$ & $\pm$ & $0.1298$ & $-0.33$ & $1.04$ \\

$D^0~~ \rightarrow K^-\pi^+$  & 
$0.0431$ & $\pm$ & $0.0376$ & $0.0805$ & $\pm$ & $0.0489$ & $-0.37$ & $0.80$ \\

$D^0~~ \rightarrow K^-\pi^+(\pi^0)$ & 
$0.0534$ & $\pm$ & $0.0405$ & $0.0936$ & $\pm$ & $0.0489$ & $-0.33$ & $0.82$ \\
\hline
~~~Average &
$0.0659$ & $\pm$ & $0.0094$ & $0.0762$ & $\pm$ & $0.0194$ & $-0.27$ & $0.54$ \\
\hline
     \end{tabular}
  \end{center}
  \caption{\label{asymmetries} Results of the two parameter fit to the
                  individual decay modes. The average centre--of--mass energy is
                  91.235\,GeV. The $\chi^2/N.D.F.$ of the averages is 0.54.}
\end{table}
\newpage
\begin{table}[h]
    \renewcommand{\arraystretch}{1.15}
  \begin{center}
    \begin{tabular}{|l|rcr|rcr|c|}
       \hline
~~~~~~~decay mode                        &
\multicolumn{3}{c|}{$A_{FB}^{c}$} &
\multicolumn{3}{c|}{$A_{FB}^{b}$} & correlation    \\
\hline
$D^{\ast+} \rightarrow (K^-\pi^+)\pi^+$   & 
$ 0.0286$ & $\pm$ & $0.1026$ & $ 0.1417$ & $\pm$ & $0.1940$ &  $-0.26$ \\

$D^{\ast+} \rightarrow (K^-\pi^+\pi^-\pi^+)\pi^+$ & 
$-0.1893$ & $\pm$ & $0.0950$ & $ 0.2492$ & $\pm$ & $0.2170$ & $-0.27$ \\

$D^{\ast+} \rightarrow (K^-\pi^+\gamma\gamma)\pi^+$ & 
$-0.2496$ & $\pm$ & $0.1187$ & $ 0.0915$ & $\pm$ & $0.2934$ & $-0.31$ \\

$D^{\ast+} \rightarrow (K^-\mu^+\nu)\pi^+$ &
$-0.0438$ & $\pm$ & $0.1656$ & $-0.3322$ & $\pm$ & $0.3417$ & $-0.38$  \\

$D^{\ast+} \rightarrow (K^-e^+\nu)\pi^+$ & 
$ 0.0342$ & $\pm$ & $0.1638$ & $ 0.4078$ & $\pm$ & $0.3746$ & $-0.23$  \\

$D^{\ast+}\rightarrow (K^-\pi^+(\pi^0))\pi^+$ & 
$-0.0514$ & $\pm$ & $0.0773$ & $ 0.0475$ & $\pm$ & $0.1763$ & $-0.27$ \\

$D^+~ \rightarrow K^-\pi^+\pi^+$  & 
$-0.0600$ & $\pm$ & $0.0903$ & $ 0.0158$ & $\pm$ & $0.3888$ & $-0.26$ \\

$D^0~~ \rightarrow K^-\pi^+$  & 
$ 0.2614$ & $\pm$ & $0.1609$ & $-0.0195$ & $\pm$ & $0.2044$ & $-0.34$ \\

$D^0~~ \rightarrow K^-\pi^+(\pi^0)$ & 
$ 0.1240$ & $\pm$ & $0.1333$ & $-0.1309$ & $\pm$ & $0.1704$ & $-0.34$ \\
\hline
~~~Average &
$-0.0496$ & $\pm$ & $0.0368$ & $ 0.0567$ & $\pm$ & $0.0756$ & $-0.28$ \\
\hline
     \end{tabular}
  \end{center}
  \caption{\label{m2asymmetries} Results of the two parameter fit to the
                  individual decay modes. The average centre--of--mass energy is
                  89.434\,GeV and the $\chi^2/N.D.F.$ of the averages is 0.96.}
\end{table}

\begin{table}[h]
    \renewcommand{\arraystretch}{1.15}
  \begin{center}
    \begin{tabular}{|l|rcr|rcr|c|}
       \hline
~~~~~~~decay mode                        &
\multicolumn{3}{c|}{$A_{FB}^{c}$} &
\multicolumn{3}{c|}{$A_{FB}^{b}$} & correlation    \\
\hline
$D^{\ast+} \rightarrow (K^-\pi^+)\pi^+$   & 
$0.1090$ & $\pm$ & $0.0772$ & $0.1214$ & $\pm$ & $0.1616$ &  $-0.25$ \\

$D^{\ast+} \rightarrow (K^-\pi^+\pi^-\pi^+)\pi^+$ & 
$0.1281$ & $\pm$ & $0.0770$ & $0.2124$ & $\pm$ & $0.1765$ & $-0.27$ \\

$D^{\ast+} \rightarrow (K^-\pi^+\gamma\gamma)\pi^+$ & 
$0.1778$ & $\pm$ & $0.1047$ & $0.1690$ & $\pm$ & $0.2535$ & $-0.29$ \\

$D^{\ast+} \rightarrow (K^-\mu^+\nu)\pi^+$ &
$0.1190$ & $\pm$ & $0.1212$ & $0.2652$ & $\pm$ & $0.2655$ & $-0.32$  \\

$D^{\ast+} \rightarrow (K^-e^+\nu)\pi^+$ & 
$0.3113$ & $\pm$ & $0.1452$ & $-0.0129$ & $\pm$ & $0.3003$ & $-0.28$  \\

$D^{\ast+}\rightarrow (K^-\pi^+(\pi^0))\pi^+$ & 
$0.1049$ & $\pm$ & $0.0720$ & $-0.0798$ & $\pm$ & $0.1494$ & $-0.25$ \\

$D^+~ \rightarrow K^-\pi^+\pi^+$  & 
$-0.0016$ & $\pm$ & $0.0902$ & $0.1387$ & $\pm$ & $0.3652$ & $-0.19$ \\

$D^0~~ \rightarrow K^-\pi^+$  & 
$0.2065$ & $\pm$ & $0.1253$ & $0.0049$ & $\pm$ & $0.1575$ & $-0.33$ \\

$D^0~~ \rightarrow K^-\pi^+(\pi^0)$ & 
$0.0384$ & $\pm$ & $0.1285$ & $0.1227$ & $\pm$ & $0.1543$ & $-0.34$ \\
\hline
~~~Average &
$0.1180$ & $\pm$ & $0.0318$ & $0.0882$ & $\pm$ & $0.0633$ & $-0.26$ \\
\hline
     \end{tabular}
  \end{center}
  \caption{\label{p2asymmetries} Results of the two parameter fit to the
                  individual decay modes. The average centre--of--mass energy is
                  92.990\,GeV and the $\chi^2/N.D.F.$ of the averages is 0.50.}
\end{table}

\begin{figure}[p]
 \begin{center}
   \mbox{\epsfig{file=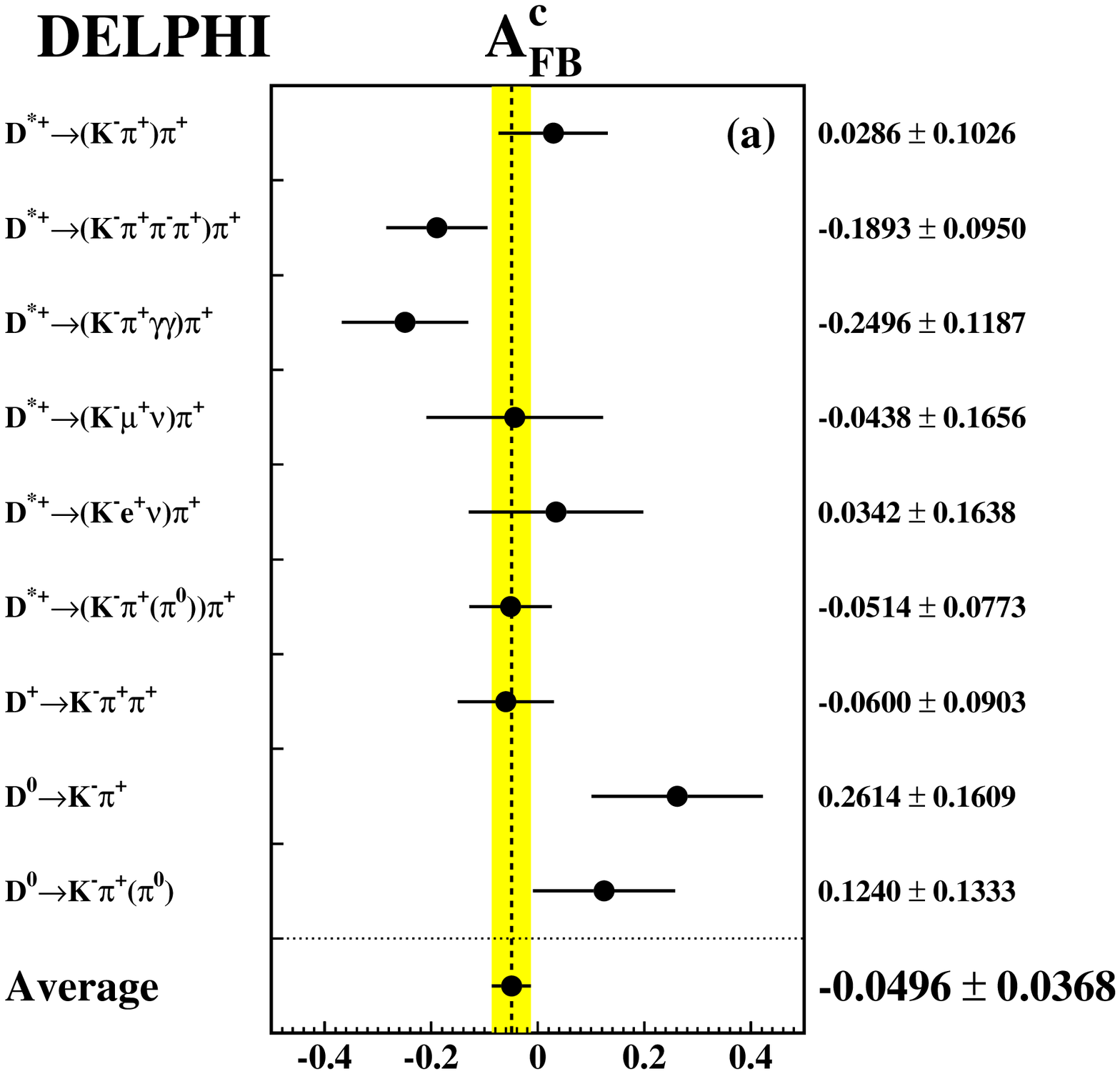,width=10cm}} \\
   \mbox{\epsfig{file=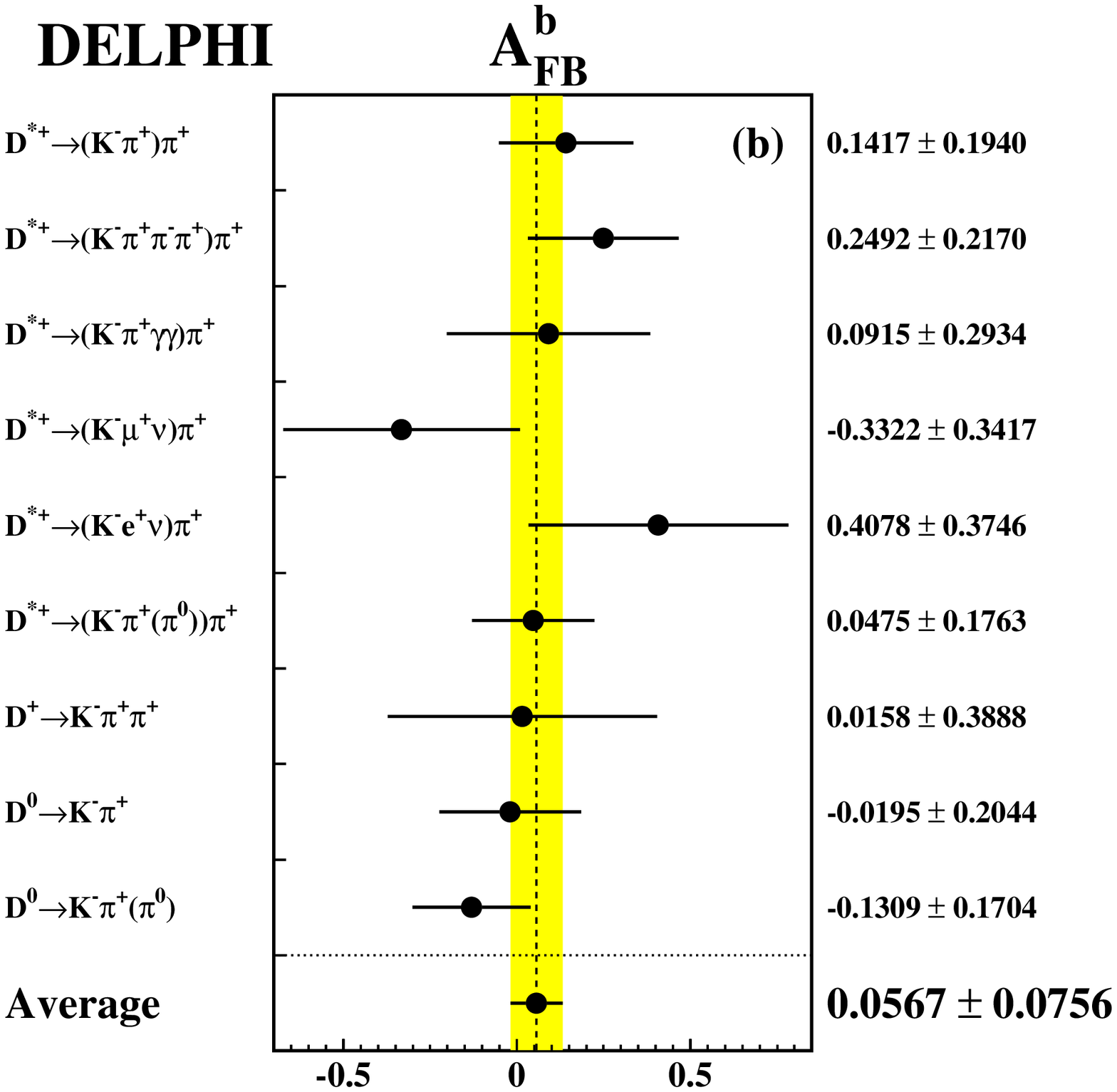,width=10cm}}
 \end{center}
 \caption[]{\label{fit_resm2} The results of the two parameter fit of 
           the $c$ and $b$ asymmetry at an average centre--of--mass
           energy of 89.434\,GeV
           for the different $D$ samples are shown.
           The grey bands represent the averages
           over all these measurements. Only statistical errors are shown.}
\end{figure}

\begin{figure}[p]
 \begin{center}
   \mbox{\epsfig{file=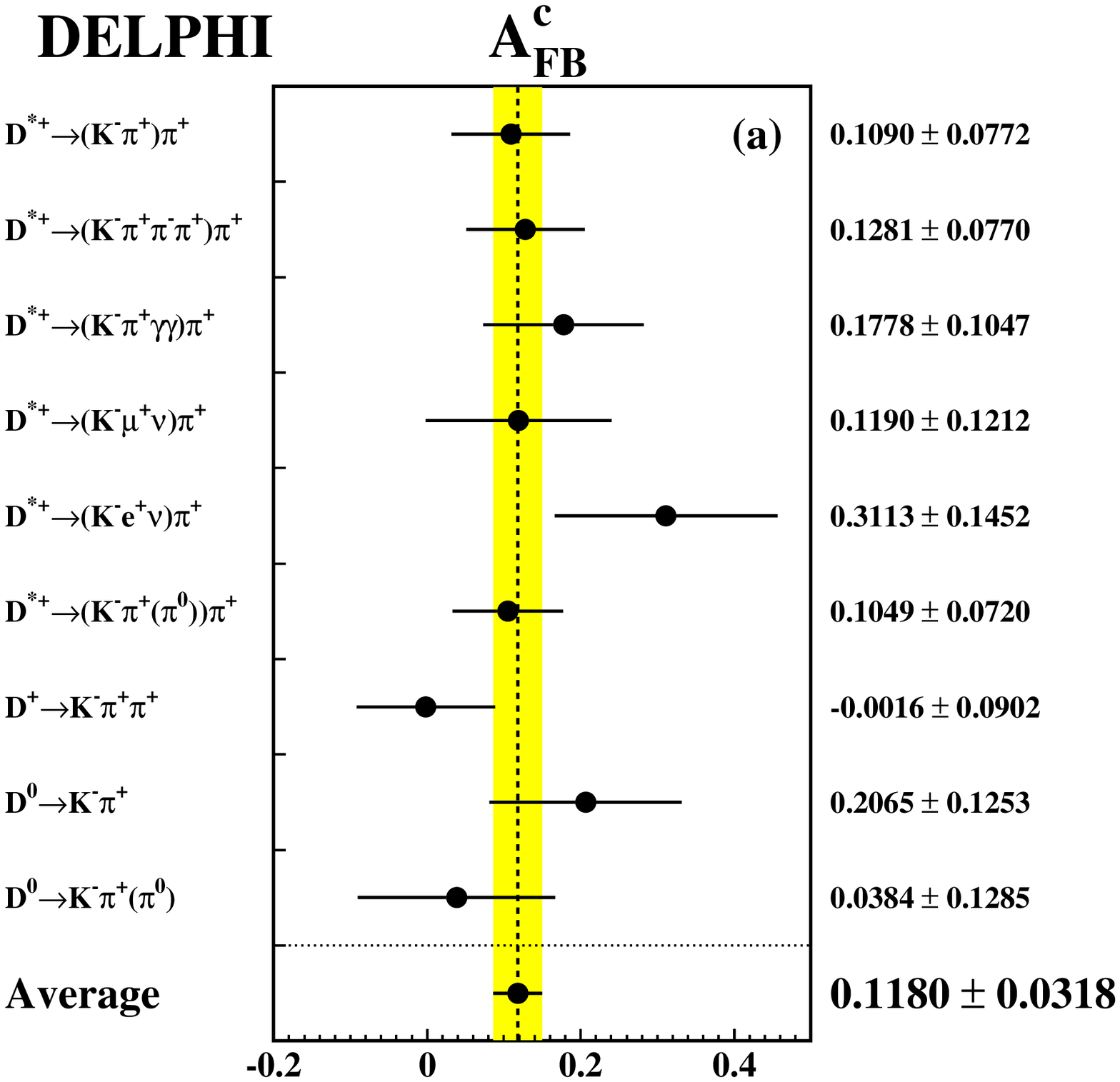,width=10cm}} \\
   \mbox{\epsfig{file=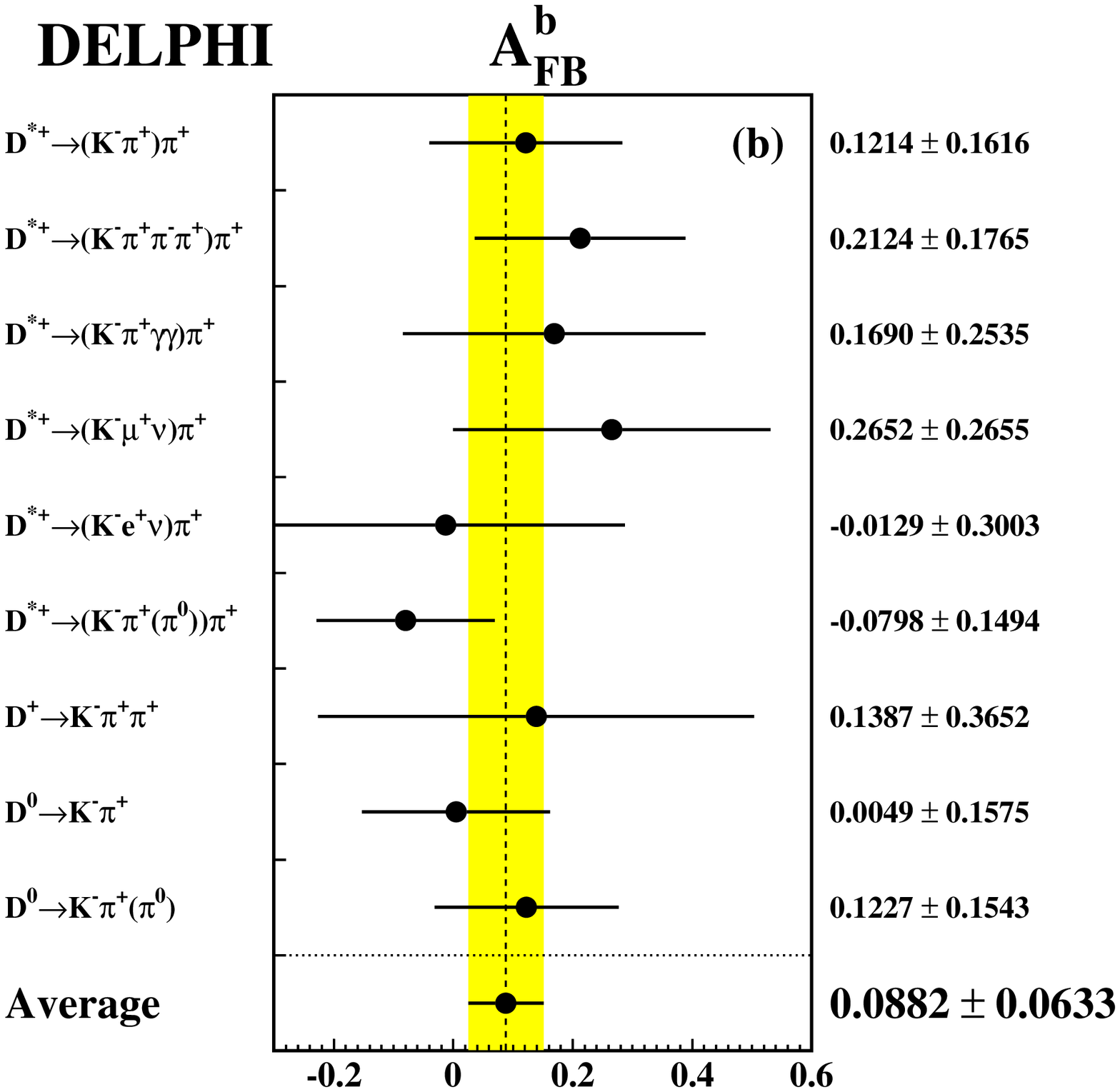,width=10cm}}
 \end{center}
 \caption[]{\label{fit_resp2} The results of the two parameter fit of 
           the $c$ and $b$ asymmetry at an average centre--of--mass
           energy of 92.99\,GeV
           for the different $D$ samples are shown.
           The grey bands represent the averages
           over all these measurements. Only statistical errors are shown.}
\end{figure}

\begin{figure}[p]
 \begin{center}
  \mbox{\epsfig{file=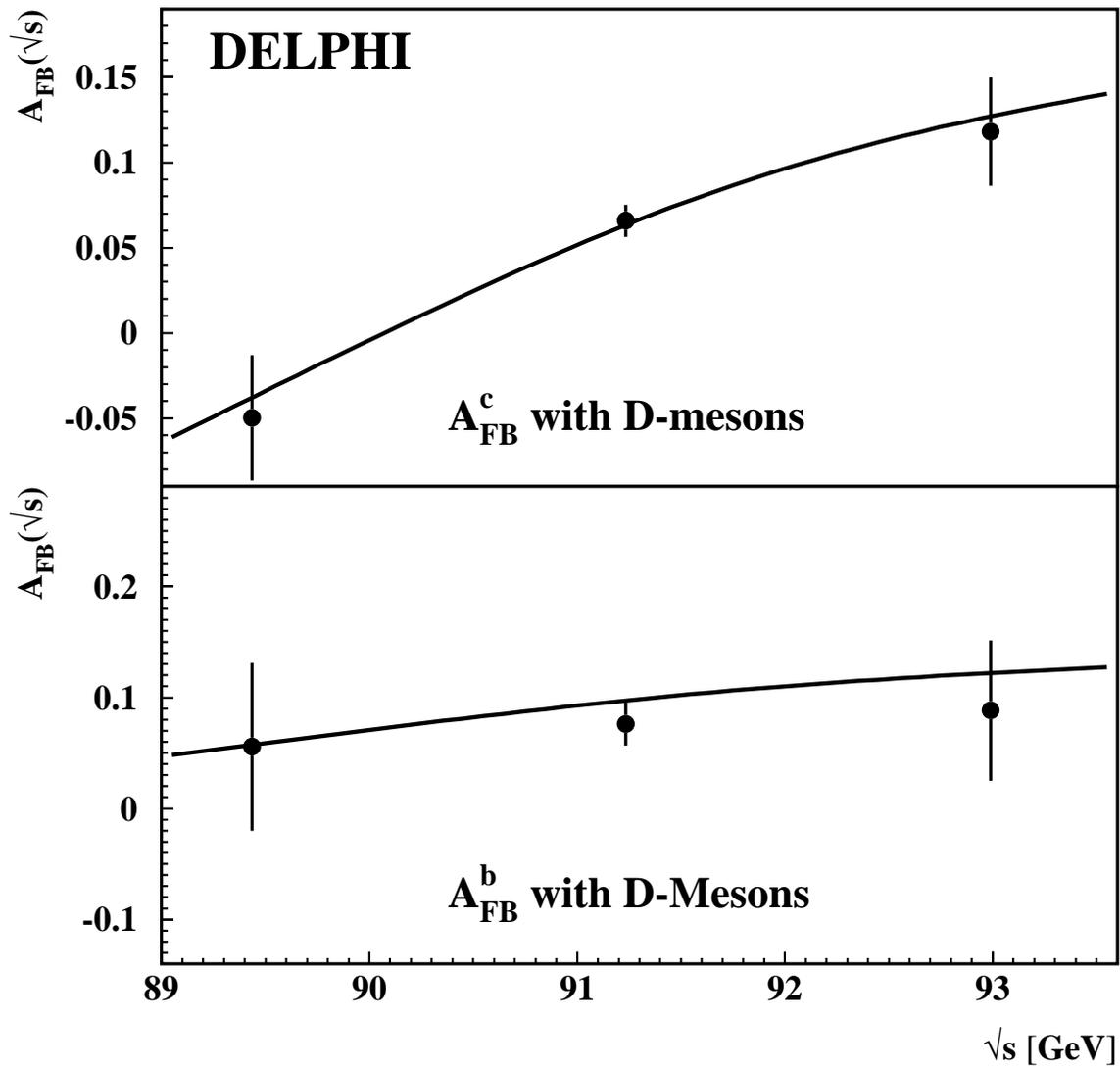,width=17cm}}
 \end{center}
\caption[]{\label{asy_energy} The $c$ and $b$ forward--backward asymmetries
           for the different average centre--of--mass energies.
           The SM prediction is also shown (M$_Z = 91.187\,{\rm GeV}/c^2$, m$_{top} =
           175.6\,{\rm GeV}/c^2$, m$_H = 300\,{\rm GeV}/c^2$, $\alpha_s({\rm M}_Z^2) = 0.120$,
           $\alpha({\rm M}_Z^2) = 1/128.896 $).}
\end{figure}

The results of the different samples at the off--peak energies are combined to
give:
\[
    \renewcommand{\arraystretch}{1.6}
\begin{array}{rcr@{\!\!\,\,}l}
   \Afbc(\sqrt{s} = 89.434\, {\rm GeV})
       &=&-&0.0496\,\, \pm\,\, 0.0368\, (stat)\\
   \Afbb(\sqrt{s} = 89.434\, {\rm GeV})
       &=& &0.0567\,\, \pm\,\, 0.0756\, (stat)\\
   \Afbc(\sqrt{s} = 92.990\, {\rm GeV})
       &=& &0.1180\,\, \pm\,\, 0.0318\, (stat)\\
   \Afbb(\sqrt{s} = 92.990\, {\rm GeV})
       &=& &0.0882\,\, \pm\,\, 0.0633\, (stat)\\
\end{array}
\]
The statistical correlation is $-0.28$ for $\sqrt{s} = 89.434$\,GeV and
$-0.26$ for $\sqrt{s} = 92.990$\,GeV. The results of the fit are shown in
figures \ref{fit_resm2} and \ref{fit_resp2}.
The averages of the $c$ and $b$ asymmetries for the different energies are 
shown in figure \ref{asy_energy}.

%=========================================================================%
\section{The effective electroweak mixing angle}
%=========================================================================%

In order to obtain the effective electroweak mixing angle from the bare 
asymmetries \Afbnc\ and \Afbnb\
at the nominal $Z$ mass, small corrections have to be applied to the measured
forward--backward asymmetries.
         The measured asymmetries for the different centre--of--mass energies have to be
         corrected to $\sqrt{s}={\rm M}_Z$. The slope of the asymmetry around ${\rm M}_Z$
         depends only on the axial vector coupling and the charge of the final
         fermions. It is thus independent of the pole asymmetry itself.
         In addition QED corrections, in particular initial state photon 
radiation, reduce the effective centre--of--mass
         energy. 
         Finally the diagrams from $\gamma$ exchange and $\gamma Z$ interference
         result in a small correction to the asymmetry.\\

The corrections have been determined using the ZFITTER program \cite{zfitter}
to be:

%\begin{table}[ht]
 \begin{center}
    \renewcommand{\arraystretch}{1.15}
  \begin{tabular}{|l||c|c|} \hline
  \multicolumn{1}{|c||}{~~~source~~~} & ~~$\delta\Afbc$~~ & ~~$\delta\Afbb$~~ \\ \hline\hline
   $\sqrt{s}={\rm M}_Z$         & $-0.0034$       &  $-0.0013$      \\ \hline
   QED effects                  & $+0.0104$       &  $+0.0041$      \\ \hline
   $\gamma$, $\gamma Z$         & $-0.0008$       &  $-0.0003$      \\ \hline\hline
   total                        & $+0.0062$       &  $+0.0025$      \\ \hline
  \end{tabular}  
  \end{center}
%\end{table}
\noindent
where the bare asymmetry is given by $\Afbnf = \Afbf + \sum_i
(\delta\Afbf)_i$. The individual QCD corrections for this analysis are
included
in the quoted measured asymmetries for $c$ and $b$. 
From the fit results for all energies, the bare asymmetries for $c$ and $b$ are
calculated to be:
$$\Afbnc=0.0715 \pm 0.0093$$
$$\Afbnb=0.0793 \pm 0.0194$$
\noindent with a total correlation of $-0.22$. The quoted error is the
combination of the statistical and systematic errors at the three different
energies.  
This leads to an effective electroweak mixing angle:
$$\SINEFF = 0.2332 \pm 0.0016\,.$$\\

%=========================================================================%
\section{Conclusion}
%=========================================================================%

A measurement of the forward--backward asymmetries \Afbc\ and \Afbb\
at \mbox{LEP 1} energies is performed using about 3.5 million hadronic $Z$ decays
collected by the DELPHI detector in the years 1992 to 1995. The heavy
quark is tagged by the exclusive reconstruction of $D$ meson decays in
the modes $\ds \to \dnpi$, $\dpl \to \kpipi$ and $\dn \to \kpipin$. The
forward--backward asymmetries for $c$ and $b$ quarks at the $Z$ resonance
are determined to be:

\[
    \renewcommand{\arraystretch}{1.6}
\begin{array}{rcl}
   \Afbc(\sqrt{s} = 91.235\,{\rm GeV}) 
        &=& 0.0659\,\, \pm\,\, 0.0094\, (stat)\,\, \pm\,\, 0.0035\, (syst) \\
   \Afbb (\sqrt{s} = 91.235\,{\rm GeV}) 
        &=& 0.0762\,\, \pm\,\, 0.0194\, (stat)\,\, \pm\,\, 0.0085\, (syst) \\
\end{array}
\]

\noindent
with a total correlation of $-0.22$. 

The analysis of the off--peak data from 1993 and 1995 leads to:
\[
    \renewcommand{\arraystretch}{1.6}
\begin{array}{rcr@{\!\!\,\,}l}
   \Afbc(\sqrt{s} = 89.434\, {\rm GeV})
       &=&-&0.0496\,\, \pm\,\, 0.0368\, (stat)\,\, \pm\,\, 0.0053\, (syst) \\
   \Afbb(\sqrt{s} = 89.434\, {\rm GeV})
       &=& &0.0567\,\, \pm\,\, 0.0756\, (stat)\,\, \pm\,\, 0.0117\, (syst) \\
   \Afbc(\sqrt{s} = 92.990\, {\rm GeV})
       &=& &0.1180\,\, \pm\,\, 0.0318\, (stat)\,\, \pm\,\, 0.0062\, (syst) \\
   \Afbb(\sqrt{s} = 92.990\, {\rm GeV})
       &=& &0.0882\,\, \pm\,\, 0.0633\, (stat)\,\, \pm\,\, 0.0122\, (syst) \\
\end{array}
\]

\noindent
with a total correlation of $-0.28$ at $\sqrt{s} = 89.434$\,GeV 
and $-0.24$ at $\sqrt{s} = 92.990$\,GeV respectively. 

The results are in good agreement with
other LEP measurements \cite{OPAL Afb D,LEP_RES} using reconstructed $D$ mesons.
The use of the full available sample of the reprocessed data for the years
1992 to 1995 leads to a significant improvement in statistical precision
as compared with the previous results from DELPHI \cite{markus}.
The results and the obtained energy dependence are consistent with the
predictions of the Standard Model.
 
From the corresponding bare asymmetries for $c$ and $b$ quarks, the effective
electroweak
mixing angle is determined as:
$$\SINEFF = 0.2332 \pm 0.0016\,.$$\\
\noindent
This result is in good agreement with the determinations of the effective
electroweak
mixing angle from several LEP and SLD measurements \cite{EWWPPE}.

\newpage
%         Created on 12-FEB-1998 by dimartino
%-------------------------------------------------------------------
\subsection*{Acknowledgements}
\vskip 3 mm
 We are greatly indebted to our technical 
collaborators, to the members of the CERN-SL Division for the excellent 
performance of the LEP collider, and to the funding agencies for their
support in building and operating the DELPHI detector.\\
We acknowledge in particular the support of \\
Austrian Federal Ministry of Science and Traffics, GZ 616.364/2-III/2a/98, \\
FNRS--FWO, Belgium,  \\
FINEP, CNPq, CAPES, FUJB and FAPERJ, Brazil, \\
Czech Ministry of Industry and Trade, GA CR 202/96/0450 and GA AVCR A1010521,\\
Danish Natural Research Council, \\
Commission of the European Communities (DG XII), \\
Direction des Sciences de la Mati$\grave{\mbox{\rm e}}$re, CEA, France, \\
Bundesministerium f$\ddot{\mbox{\rm u}}$r Bildung, Wissenschaft, Forschung 
und Technologie, Germany,\\
General Secretariat for Research and Technology, Greece, \\
National Science Foundation (NWO) and Foundation for Research on Matter (FOM),
The Netherlands, \\
Norwegian Research Council,  \\
State Committee for Scientific Research, Poland, 2P03B06015, 2P03B03311 and
SPUB/P03/178/98, \\
JNICT--Junta Nacional de Investiga\c{c}\~{a}o Cient\'{\i}fica 
e Tecnol$\acute{\mbox{\rm o}}$gica, Portugal, \\
Vedecka grantova agentura MS SR, Slovakia, Nr. 95/5195/134, \\
Ministry of Science and Technology of the Republic of Slovenia, \\
CICYT, Spain, AEN96--1661 and AEN96-1681,  \\
The Swedish Natural Science Research Council,      \\
Particle Physics and Astronomy Research Council, UK, \\
Department of Energy, USA, DE--FG02--94ER40817. \\
%=========================================================================%
\newpage
%=========================================================================%

%=========================================================================%
% now the figures
%=========================================================================%
\parindent0pt
\newpage

\newpage
%%
%%---- PAGE FOUR -------- figure 4 ---------------------------------------------
%                                  (fit_1)
\newpage

\newpage

\newpage

\newpage

\newpage

\newpage

%%% Local Variables: 
%%% mode: latex
%%% TeX-master: "paper"
%%% End: 

\end{document}